\documentclass[a4paper,11pt]{article}
\pdfoutput=1 
\usepackage{jcappub}
\usepackage[T1]{fontenc} 

\usepackage[utf8]{inputenc}
\usepackage{graphicx}
\usepackage{booktabs}
\usepackage{tabularx}
\usepackage{textcomp}
\usepackage{xspace}
\usepackage{amsmath}
\usepackage{mathtools}
\usepackage{amsfonts}
\usepackage{amssymb}
\usepackage{tablefootnote}
\usepackage[dvipsnames,table]{xcolor}
\usepackage{mathrsfs}
\usepackage[normalem]{ulem}
\usepackage{mathrsfs}

\hypersetup{urlcolor=BlueViolet,
	    citecolor=Plum,
	    linkcolor=PineGreen}
\usepackage[official]{eurosym}
\definecolor{lightgray}{rgb}{0.9,0.9,0.9}	    
\definecolor{green}{rgb}{0,0.5,0}
\definecolor{red}{rgb}{1,0,0}
\definecolor{blue}{rgb}{0,0,0.5}

\newcommand\codo[1]{{\tt #1}}

\def\Plus{\raisebox{-0.2ex}{\large\texttt{+}}}

\title{One trick to treat them all: SuperEasy linear response for any hot dark matter in $N$-body simulations}

\author[a]{Giovanni Pierobon,}
\author[b,c]{Markus R. Mosbech,}
\author[d,a]{Amol Upadhye,}
\author[a]{Yvonne Y.~Y.~Wong}

\affiliation[a]{Sydney Consortium for Particle Physics and Cosmology, School of Physics, University of New South Wales, Sydney NSW 2052, Australia}
\affiliation[b]{Institute for Theoretical Particle Physics and Cosmology (TTK), RWTH Aachen University, D-52056 Aachen, Germany}
\affiliation[c]{Institute for Theoretical Particle Physics (TTP), Karlsruhe Institute of Technology (KIT), 76128 Karlsruhe, Germany}
\affiliation[d]{South-Western Institute for Astronomy Research, Yunnan University, Kunming 650500, P.R. China}

\emailAdd{g.pierobon@unsw.edu.au}
\emailAdd{mosbech@physik.rwth-aachen.de}
\emailAdd{a.upadhye@ynu.edu.cn}
\emailAdd{yvonne.y.wong@unsw.edu.au}

\abstract{We generalise the SuperEasy linear response method, originally developed to describe massive neutrinos in cosmological $N$-body simulations,
to any hot dark matter (HDM) species with arbitrary momentum distributions.  The method uses analytical solutions of the HDM phase space perturbations in various limits and constructs from them a modification factor to the gravitational potential that tricks the cold particles into trajectories as if HDM particles were present in the simulation box.
The modification factor is algebraic in the cosmological parameters and requires no fitting.  
Implementing the method in a Particle-Mesh simulation code and testing it on HDM cosmologies up to the equivalent effect of $\sum m_\nu = 0.315$~eV-mass neutrinos, we find that the generalised SuperEasy approach is able to predict the total matter and cold matter power spectra to $\lesssim 0.1\%$ relative to other linear response methods and  to $\lesssim 0.25\%$ relative to particle HDM simulations.  Applying the method to cosmologies with mixed neutrinos\Plus{}thermal QCD axions and neutrinos\Plus{}generic thermal bosons, we find that non-standard HDM cosmologies have no intrinsically different non-linear signature in the total matter power spectrum from standard neutrino cosmologies.   However, because they predict different time dependencies even at the linear level and the differences are augmented by non-linear evolution, it remains a possibility that observations at multiple redshifts may help distinguish between them.}

\begin{document}
\begin{flushright}
	{\large \tt CPPC-2024-08}\\
        {\large \tt TTK-24-35}\\
        {\large \tt TTP24-36}
\end{flushright}	
	
\maketitle
\flushbottom

\section{Introduction}

Cosmologists have long known that precision cosmological observations can constrain the density fraction of hot but non-relativistic neutrinos in the universe~\cite{Hu:1997mj}.  The standard model (SM) of particle physics also makes a clear prediction of the properties of these neutrinos~\cite{Bennett:2019ewm,Akita:2020szl,Froustey:2020mcq,Bennett:2020zkv,Drewes:2024wbw} up to their absolute masses.  
Constraining this basic scenario---and from which the absolute neutrino mass scale---has become a staple of modern cosmological analyses, from probes of big bang nucleosynthesis (BBN)~\cite{Fields:2019pfx,Yeh:2022heq} and the cosmic microwave background (CMB)~\cite{Planck:2018vyg}, to late-universe observables such as weak gravitational lensing~\cite{DES:2022ccp} and galaxy redshift surveys~\cite{Ivanov:2019hqk}.  Assuming a minimal model of the dark energy, i.e., the cosmological constant, combinations of these observations have constrained the sum of neutrino masses to $\sum m_\nu \lesssim 0.12$~eV~\cite{Planck:2018vyg,DESI:2024mwx}.  Over the next several years, cosmology is expected to go beyond upper bounds to achieve a measurement of the neutrino mass sum~\cite{EuclidTheoryWorkingGroup:2012gxx,CMB-S4:2016ple}.

We are thus entering an era in which cosmological data will have sufficient precision to allow us to challenge the standard prediction itself. Indeed, physically well-motivated non-standard hot dark matter (HDM) scenarios abound.  To name but a few, neutrinos may have non-standard interactions with dark matter~\cite{Wilkinson:2014ksa,Mosbech:2020ahp} or dark energy~\cite{Bi:2004ns}, altering their distribution function and their late-universe clustering;  One or more light ``sterile'' states in addition to the three SM neutrinos might explain persistent anomalies in short-baseline neutrino experiments and also leave observable cosmological signatures~\cite{Acero:2022wqg}; The axion~\cite{Peccei:1977hh}, a light pseudoscalar particle proposed as a solution to the strong $CP$ problem, may couple strongly enough to SM particles to attain thermal equilibrium  after inflation, making the axion a HDM similar to neutrinos~\cite{Hannestad:2010yi,Archidiacono:2013cha,Marsh:2015xka,Ferreira:2018vjj,DEramo:2021psx,DEramo:2021lgb,Notari:2022ffe,DEramo:2022nvb, Bianchini:2023ubu}.  Some of these scenarios have been proposed as solutions to decade-old tensions between different cosmological probes~\cite{DEramo:2018vss,Sakstein:2019fmf,Duan:2021whx,Das:2021pof,Sgier:2021bzf}.  Others, including the idea of non-thermal neutrino distributions~\cite{Oldengott:2019lke,Alvey:2021sji}, have been explored in the recent literature as a way to relax cosmological neutrino mass bounds.  

While the same broad-brush principles apply to constraining these non-standard HDM scenarios cosmologically as they do to standard neutrino masses, their detailed signatures differ, which raises the possibility that these scenarios could eventually be distinguished from each other and also from the standard neutrino HDM scenario itself.  In the linear regime of perturbation evolution, model-specific signatures can be straightforwardly modelled following standard cosmological perturbation theory~\cite{Ma:1995ey} and numerically solved using a linear Boltzmann code such as  \codo{camb}~\cite{Lewis:1999bs,Howlett:2012mh} or  \codo{class}~\cite{Blas:2011rf,Lesgourgues:2011rh}.  As forthcoming large-scale structure observations move into the non-linear regime, however, the breakdown of linear perturbation theory not only invokes the prospect of new HDM signatures on non-linear scales, but also precipitates the need to examine how to model these effects reliably. \cite{Font-Ribera:2013rwa,Elbers:2020lbn,Angulo:2021kes}

On non-linear scales, the main difficulty in predicting HDM signatures lies in modelling the gravitational clustering of HDM particles in an accurate way.  Unlike photons, HDM can make up a non-negligible fraction of the universe's energy density at late times and cluster significantly on large scales like cold dark matter (CDM) and baryons.  Unlike CDM and baryons, however, HDM particles often have thermal velocities comparable to, or even greater than, the velocities they acquire via infall into the gravitational potential wells of dark matter structures.  This last fact necessitates that we track the full six-dimensional phase space of HDM particles, rather than merely their three-dimensional spatial distribution as we do for CDM, in the computation of HDM signatures in the large-scale structure distribution.

The gold standard for the investigation of large-scale structure in the non-linear regime is the $N$-body simulation,
wherein a cold matter (i.e., CDM+baryon) population is represented by a set of particles moving under their mutual gravitational forces (see, e.g.,~\cite{Angulo:2021kes} for a recent review).  These simulations are however computationally expensive---and become even more so when HDM particles are added to the mix (e.g.,~\cite{Viel:2010bn,Banerjee:2016zaa,Yu:2016yfe,Emberson:2016ecv,Inman:2016prk,Villaescusa-Navarro:2019bje,Elbers:2020lbn,Parimbelli:2022pmr,Hernandez-Aguayo:2024slb}).%
\footnote{Alternatively, instead of representing HDM with particles, one could solve the full six-dimensional Vlasov equation for the HDM phase space density~\cite{Yoshikawa:2020ehd}.}
A convenient way to circumvent this issue that has emerged in the past decade is to adopt a split approach, which broadly consists of employing a linear response theory to track the HDM perturbations on a Particle-Mesh (PM) grid, while retaining the cold matter population, which supplies the external potential to which the HDM population responds, on a full particle representation~\cite{Ali-Haimoud:2012fzp,Dakin:2017idt}.  Such a split, linear response approach is generally unable to capture to full non-linear evolution of the HDM component.  However, for a HDM population that does not cluster too strongly---for SM neutrinos this normally means neutrino masses satisfying $\sum m_\nu \lesssim 1$~eV~\cite{Brandbyge:2008js,Bird:2018all,Chen:2022dsv}---the approach should be sufficient to compute the cold matter power spectrum to percent-level accuracy.
Several linear response methods/implementations with varying degrees of complexity and accuracy have been proposed for SM neutrinos~\cite{{Liu:2017now},Fidler:2018bkg,Partmann:2020qzb,Heuschling:2022rae}, including the SuperEasy~\cite{Chen:2020kxi} and multi-fluid Linear Response (MFLR)~\cite{Chen:2020bdf} methods developed by some of us.  For non-standard HDM candidates, however, the non-linear aspects of their perturbation evolution remain to be investigated in detail, from both the perspectives of their signatures and the methodology with which to compute them.

The purpose of the present work is therefore two-fold: (i)~to generalise the SuperEasy linear response method to any HDM, and (ii)~to use the method to investigate the particular signatures of non-standard HDM on non-linear scales.  In short, the SuperEasy method consists in finding asymptotic solutions to the {\it linearised} Vlasov-Poisson system of equations for the density perturbations of a subdominant HDM in both the clustering limit on large length scales and the free-streaming limit on small scales. For SM neutrinos, the limiting solutions are analytical and algebraic in the wave number~$k$, the matter density $\Omega_{\rm m}$, the individual neutrino masses $m_\nu$, and the scale factor~$a$.  A rational function can be used to interpolate between the two limits~\cite{Ringwald:2004np,Wong:2008ws}, resulting in a simple multiplicative, $k$-dependent algebraic modification to the gravitational source term in the Poisson equation that is able to capture the effect of SM neutrino masses on the evolution of the cold matter.  The method is easy to implement in any simulation code that uses a PM gravity solver, and incurs virtually no runtime overhead relative to a standard cold matter-only simulation~\cite{Chen:2020kxi}.

Generalising the SuperEasy method beyond SM neutrinos complicates the scheme somewhat.  This is because once the HDM background phase space distribution deviates from a relativistic Fermi-Dirac form, we are no longer assured of either the existence of analytical limiting solutions or even if these limiting solutions are finite and well-defined.  Specifically, the key quantity in the free-streaming solution is the momentum-averaged free-streaming scale $k_{\rm FS} \propto \sqrt{1/\langle p^2\rangle}$, which even for the seemingly innocuous relativistic Bose-Einstein distribution turns out to be formally infinite. However, as explored in reference~\cite{Chen:2020kxi}, asymptotic solutions and hence interpolations can nonetheless be constructed at the momentum-by-momentum level.  This opens up a new way to apply the Super-Easy approach to modelling virtually all conceivable subdominant HDM.  Finally, we note that while we focus our attention on non-linear evolution, we anticipate that our method can work also in the context of linear Boltzmann codes like \codo{camb} and \codo{class} at low redshifts to speed up calculations.

The article proceeds as follows. Section~\ref{sec:generalised_linear_response_model}
introduces the linearised non-relativistic Vlasov-Poisson system that describes the linear response of a generic HDM species to an external potential.  We construct the SuperEasy linear response method in section~\ref{sec:gse} applicable to any HDM.  This method is then applied 
in section~\ref{sec:n-body_simulations} to $N$-body simulations of several mixed HDM cosmologies and 
contrasted section~\ref{sec:comparison} with the outcomes of (i) other linear response simulations and (ii) particle-based simulations.  Section \ref{sec:conclusions} contains our conclusions.


\section{Linear response of a subdominant hot dark matter}
\label{sec:generalised_linear_response_model}

The starting point of our linear response approach for a generic subdominant HDM is the linearised Vlasov (also called collisionless Boltzmann) equation,
\begin{equation}
	\frac{\mathrm{d} \delta f}{\mathrm{d}s} \equiv \frac{\partial \delta f}{\partial s} + \frac{\vec{p}}{m} \cdot \nabla_{\!\vec{x}} \, \delta f - a^2m \nabla_{\!\vec{x}} \, \Phi \cdot \nabla_{\!\vec{p}} \, \bar{f} = 0,  
	\label{Eq:BoltzmannEq}
\end{equation}
which, in this form, describes the time evolution of the phase space density contrast $\delta f=\delta f(\vec{x},\vec{p},s)\equiv f(\vec{x},\vec{p},s)-\bar{f}(p)$ of a hot dark matter species of particle mass~$m$ in the presence of a Newtonian gravitational potential  $\Phi = \Phi(\vec{x},s)$
in an expanding background.  Here, $\vec{x}$ is the comoving spatial coordinate, 
$s \equiv \int a^{-2} \mathrm{d}t$ is the superconformal time,  
 $\vec{p}$ the comoving momentum, and $a=a(s)$ the scale factor.   The equation is linearised in that the last term contains the $x$-independent $\nabla_{\!\vec{p}} \, \bar{f}$ instead of the more general $\nabla_{\!\vec{p}} \, f$ (which necessitates a convolution product).
 We do not yet specify a particular form for the homogeneous background phase space distribution $\bar{f}(p)$, but note that after decoupling  $\bar{f}(p)$ is independent of time.

The gravitational potential~$\Phi$ is sourced by fluctuations in the total matter density according to  the Poisson equation,
 \begin{equation}
 \nabla^2_{\vec x} \, \Phi(\vec{x},s) 
 = \frac{3}{2} {\cal H}^2(s) \, \Omega_{\rm m}(s)  \, \delta_{\rm m} (\vec{x},s),
 \label{eq:poisson}
 \end{equation}
where ${\cal H} \equiv aH $ is the conformal Hubble expansion rate, $\Omega_{\rm m}(s) \equiv \bar{\rho}_{\rm m}(s)/\rho_{\rm crit}(s)$  the time-dependent reduced matter density, and we take the total matter density $\bar{\rho}_{\rm m} = \bar{\rho}_{\rm cb}+ \bar{\rho}_{\rm hdm}$ to comprise a dominant cold component (subscript ``cb'' for cold dark matter and baryons) and a subdominant hot component (subscript ``hdm'').  It follows that 
\begin{equation}
\delta_{\rm m} \equiv f_{\rm cb} \delta_{\rm cb} + f_{\rm hdm} \delta_{\rm hdm}
\end{equation}
defines the total matter density contrast, where the hot component's density contrast
can  be constructed from the solution to equation~\eqref{Eq:BoltzmannEq} via
\begin{equation}
\delta_{\rm hdm}  (\vec{x},s) \equiv \frac{\rho_{\rm hdm}(\vec{x},s)-\bar{\rho}_{\rm hdm}(s)}{\bar{\rho}_{\rm hdm}(s)} = \frac{\int {\rm d}^3 p\, \delta f(\vec{x},\vec{p},s)}{\int {\rm d}^3 p \, \bar{f}(p)},
\label{eq:nudensitycontrast}
\end{equation} 
and $f_{\rm cb}  \equiv \bar{\rho}_{\rm cb}/\bar{\rho}_{\rm m}$, $f_{\rm hdm}  \equiv \bar{\rho}_{\rm hdm}/\bar{\rho}_{\rm m}$ are time-independent weights.

To implement linear response, we switch to Fourier space. Defining the Fourier transform to be $A(\vec{k}) = \mathscr{F}[A(\vec{x})] \equiv \int_{-\infty}^{\infty} A(\vec{x}) \, e^{-i \vec{k} \cdot \vec{x}} \, \mathrm{d}^3x$, equations~\eqref{Eq:BoltzmannEq} and~\eqref{eq:poisson} can be recast respectively into~\cite{Bertschinger:1993xt}
\begin{equation}
	\frac{\partial {\delta f}}{\partial s} + \frac{i \vec{k} \cdot \vec{p}}{m} {\delta f}-  i m a^2 \vec{k} \, {\Phi} \cdot \nabla_{\!\vec{p}}  \bar{f}  = 0 \, ,
	\label{Eq:FourierBoltzmann}
\end{equation}
and 
\begin{equation}
 k^2 \, \Phi(\vec{k},s) 
 = -\frac{3}{2} {\cal H}^2(s) \, \Omega_{\rm m}(s)  \, \delta_{\rm m} (\vec{k},s).
 \label{eq:poissonk}
\end{equation}
Given that the gravitational potential is sourced predominantly by cold matter, we could effectively treat ${\Phi}(\vec{k},s)$ as an external potential. 
 Then, equation~\eqref{Eq:FourierBoltzmann} admits a formal solution
\begin{equation}
	\begin{aligned}
		{\delta f}(\vec{k},\vec{p},s) =& \, {\delta f}(\vec{k},\vec{p},s_{\rm i}) \, \text{exp} \! \left(-\frac{i\vec{k}\cdot\vec{p}}{m} (s-s_{\rm i})\right) \\
		& \hspace{1.5cm} + i m \vec{k} \cdot \nabla_{\!\vec{p}} \, \bar{f} \int_{s_{\rm i}}^s \mathrm{d}s' \, a^2(s') \, {\Phi}(\vec{k},s') \, \text{exp} \! \left(-\frac{i\vec{k}\cdot\vec{p}}{m} (s-s') \right) \, ,
	\end{aligned}
	\label{Eq:GenSolution}
\end{equation}
where the subscript ``i'' denotes the initial state.  As in \cite{Chen:2020kxi}, neglecting at late times (i.e., $s \gg s_{\rm i}$) the initial space density contrast ${\delta f}(\vec{k},\vec{p},s_{\rm i}) \ll {\delta f}(\vec{k},\vec{p},s)$  yields
\begin{equation}
		{\delta f}(\vec{k},\vec{p},s) = i m \vec{k} \cdot \nabla_{\!\vec{p}} \, \bar{f} \int_{s_{\rm i}}^s \mathrm{d}s' \, a^2(s') \, {\Phi}(\vec{k},s') \, \text{exp} \! \left(-\frac{i\vec{k}\cdot\vec{p}}{m} (s-s') \right) \, .
	\label{Eq:GenSolution2}
\end{equation}
This is the linear response solution at the level of the individual HDM momentum $\vec{p}$.
It is possible to further integrate this expression in $\vec{p}$ to yield a linear response solution for the total HDM density contrast $\delta_{\rm hdm}(\vec{k},s)$ itself---this has been done many times in, e.g., \cite{Ringwald:2004np,Chen:2020kxi}, for SM neutrinos.  For the purpose of this work, however, we shall defer the $p$-integration to a later stage in our calculation.


\section{Generalised SuperEasy linear response}\label{sec:gse}

The strategy of the generalised SuperEasy linear response method is to manipulate the linear response solution~\eqref{Eq:GenSolution2} into a simple algebraic form.  To do so, we first average the expression~\eqref{Eq:GenSolution2} over $\mu\equiv \hat{k}\cdot\hat{p}$, i.e.,
$\langle \delta f\rangle_\mu  (\vec{k},p,s) \equiv (1/2) \int_{-1}^{1} {\rm d} \mu\, \delta f(\vec{k},\vec{p},s)$, to obtain%
\footnote{We note that this averaging procedure is {\it not} equivalent to averaging $\Phi(\vec{k},s)$ over the direction $\hat{k}$.}
\begin{equation}
	\left< \delta f \right>_\mu (\vec{k}, p,s)
	 =
	m k \frac{\partial \bar f}{\partial p}
	\int_{s_{\rm i}}^s \mathrm{d}s' a^2(s') \Phi(\vec{k},s') {\cal W}\left[\frac{kp(s-s')}{m}\right],
 	\label{eq:df_sprime}
\end{equation}
with ${\cal W}(q)=\sin(q)/q^2 - \cos(q)/q$. For a late-time $\Lambda$CDM universe, equation~\eqref{eq:df_sprime} can also be written in the form
\begin{equation}
\begin{aligned}
 \left< \delta f \right>_\mu (\vec{k}, p,s)
 =\, & - \sqrt{\frac{3}{2}}  \frac{\partial \bar{f}}{\partial \ln p} \frac{k_{{\rm FS},p}(s)}{k} \left(\frac{a_\Lambda}{a(s)}\right)^{3/2} \frac{\delta_{\rm m} (\vec{k},s)}{g(\vec{k},s)} \\
 & \times \int_{a_{\rm i}/a_\Lambda}^{a(s)/a_\Lambda} \! {\rm d}y\,  \frac{y \,  g(\vec{k},s'(y)) }{\sqrt{1+y^3}}\, {\cal W} \left[\frac{k}{k_{{\rm FS},p}(s)} \sqrt{\frac{a(s)}{a_\Lambda}} \int^{a(s)/a_\Lambda}_{a'/a_\Lambda} \! \frac{{\rm d}y'}{\sqrt{y'^3(1+y'^3)}}\right],
	\label{eq:df_sprimeLCDM}
 \end{aligned}
\end{equation}
where $g(\vec{k},s)=D(\vec{k},s)/a(s)$ is the reduced linear growth factor normalised to $g=1$ in the matter-domination era if the matter is entirely cold,
$k_{{\rm FS},p}$ is the $p$-dependent free-streaming scale defined as%
\footnote{The definition of $k_{{\rm FS},p}$ here differs from that in reference~\cite{Chen:2020kxi} by a factor $\sqrt{3}$, but is consistent with the definition in reference~\cite{Chen:2020bdf}.}
\begin{equation}
    k_{{\rm FS},p}(s)\equiv \sqrt{\frac{3}{2}}\frac{m a(s)}{p}\mathcal{H}(s)\Omega_{\rm  m}^{1/2}(s)= \sqrt{\frac{3}{2}} \,\frac{m}{p} H_0 \sqrt{a(s)\, \Omega_{\rm m,0}}\, ,
\end{equation}
and $a_\Lambda$ corresponds to the epoch of matter-$\Lambda$ equality.

\begin{figure}[t]
\centering
\includegraphics[width=\textwidth]{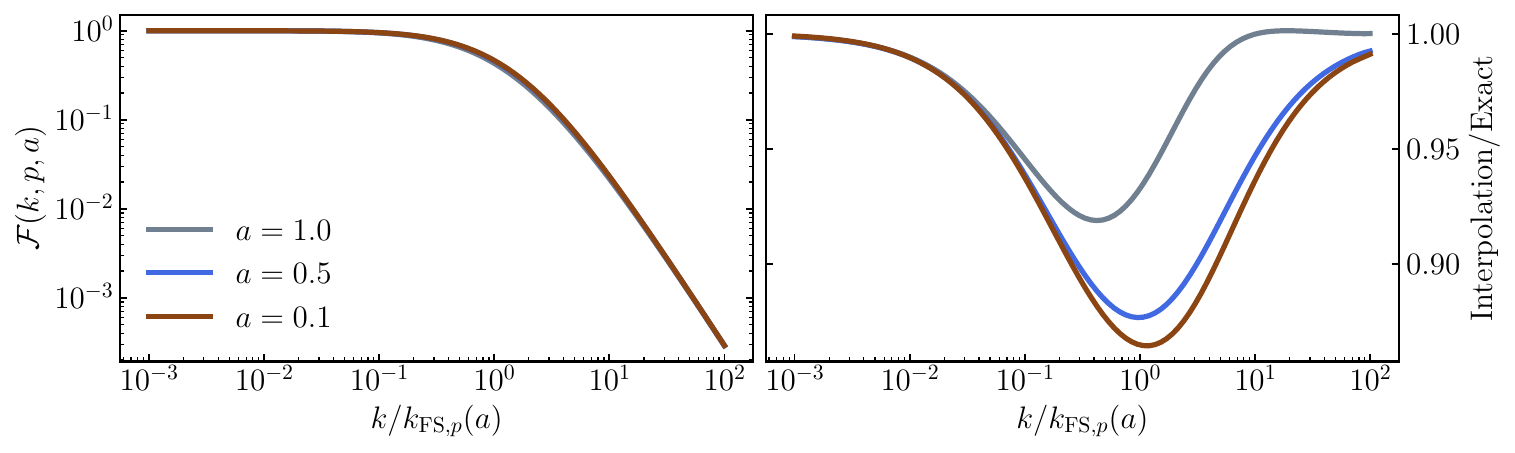}
\caption{{\it Left}: Exact solution of equation~\eqref{eq:df_sprime} at $a=1, 0.5, 0.1$ (grey, blue, brown), expressed in terms of the function ${\cal F}(k,p,s)$ defined in equation~\eqref{eq:dfmu}. We have assumed an external potential of the form $\Phi(s) = g(s) \Phi(s_{\rm i})$, where $g(s)$ is the standard $\Lambda$CDM growth factor for a reference cosmology with  $\omega_{\rm m}=0.1424$ and $h=0.6766$. 
{\it Right}: The interpolation function~\eqref{eq:interpolationf} normalised to the exact solution at the same set of scale factors.}
\label{fig:interpolationf}
\end{figure}

Expressing equation~\eqref{eq:df_sprime} (and hence~\eqref{eq:df_sprimeLCDM}) in the form
\begin{equation}
     \langle\delta f\rangle_{\mu} (\vec{k},p,s)  = -\frac{1}{3}\frac{\partial \bar{f}}{\partial \ln p} {\cal F}(k,p,s) \delta_{\rm m}(\vec{k},s) ,
     \label{eq:dfmu}
\end{equation} 
reference~\cite{Chen:2020kxi} showed that the function ${\cal F}(k,p,s)$ has the asymptotic behaviours 
\begin{equation}
{\cal F}(k/k_{{\rm FS},p}\to 0) \to 1,   \qquad
{\cal F}(k/k_{{\rm FS},p}\to \infty) \to  3\, \frac{k_{{\rm FS},p}^2}{k^2}.
\label{eq:asymptotic}
\end{equation}
In the region between $k/k_{{\rm FS},p} \to 0$ and $k/k_{{\rm FS},p}\to \infty$, no simple analytical solution exists for ${\cal F}(k,p,s)$. Reference~\cite{Chen:2020kxi} used for simplicity an algebraic interpolation function
\begin{equation}
{\cal F}(k,p,s)=\frac{3\, k_{{\rm FS},p}^2}{(k+\sqrt{3} \, k_{{\rm FS},p})^2}
\label{eq:interpolationf}
\end{equation}
to connect the two limits.   Figure~\ref{fig:interpolationf} shows this interpolation function juxtaposed with the exact solution assuming a $k$-independent external potential given by $\Phi(s) = g(s) \Phi(s_{\rm i})$, where $g(s(a)) = (5/2) (H(a)/a) \, \Omega_{\rm m,0} \int_0^a {\rm d} a' /(a' H(a'))^3$ is the $\Lambda$CDM linear growth factor.%
\footnote{Of course, in a realistic situation $\Phi$ is not external and receives feedback from the HDM density fluctuations themselves. A consistent linear growth factor therefore must depend on $k$.  This $k$-dependence can be expected to impact on the rate of change of $g(k,s)$ at order $\Omega_{\rm hdm}/\Omega_{\rm m}$ around the free-streaming scale $k_{{\rm FS},p}$.}
We note however any other monotonically decreasing function in $k/k_{{\rm FS},p}$ that reproduces the correct asymptotic behaviours will work to varying degrees of accuracy.

Using equation~\eqref{eq:dfmu} we can construct a linear response solution for the total HDM density contrast by a simple momentum integration, 
\begin{equation}
    \delta_{\rm hdm}(\vec{k},s)=\frac{1}{C}\int_0^\infty {\rm d}p\, p^2 \ \langle\delta f\rangle_{\mu}(\vec{k},p,s)
    = - \frac{1}{3C}\left[\int_0^{\infty} {\rm d} p \, p^3\frac{\partial \bar{f}}{\partial p} {\cal F}(k,p,s) \right] \delta_{\rm m}(\vec{k},s),
\label{eq:dh1}        
\end{equation}
where $C\equiv \int_0^\infty {\rm d} p\, p^2\ \bar{f}(p)$.  With a suitable choice of interpolation function ${\cal F}(k,p,s)$, equation~\eqref{eq:dh1} suffices in principle as a generalised SuperEasy linear response solution of $\delta_{\rm hdm}(\vec{k},s)$.  In practice, however, we find it convenient to further manipulate the expression by way of integration by parts, to obtain
\begin{equation}
    \delta_{\rm hdm}(\vec{k},s) = \frac{1}{C}\left[\int_0^{\infty} {\rm d} p \, p^2 \bar{f} (p) \, {\cal G}(k,p,s) \right]\delta_{\rm m}(\vec{k},s),
\label{eq:dh2}        
\end{equation}
with 
\begin{equation}
{\cal G}(k,p,s) \equiv {\cal F}(k,p,s) +
 \frac{1}{3} \frac{\partial {\cal F}}{\partial \ln p}.
 \label{eq:FtoG}
\end{equation}
In comparison with~\eqref{eq:dh1}, the new linear response solution~\eqref{eq:dh2} has the advantage that the quantity ${\rm d}p\,p^2 \bar{f}/C$ has a clear physical interpretation: it is the fractional HDM number density in the momentum bin $(p, p+{\rm d} p)$. In fact, this last step allows us to formally connect the standard, Eulerian formulation of cosmological perturbations (represented by the Vlasov equation) to the semi-Lagrangian multi-fluid linear response (MFLR) approach of~\cite{Chen:2020bdf}, where the relation~\eqref{eq:FtoG} provides the transformation between the solutions of the two approaches.  See appendix~\ref{app:SuperEasytoMFLR} for more details on this correspondence.

\begin{figure}[t]
\centering
\includegraphics[width=\textwidth]{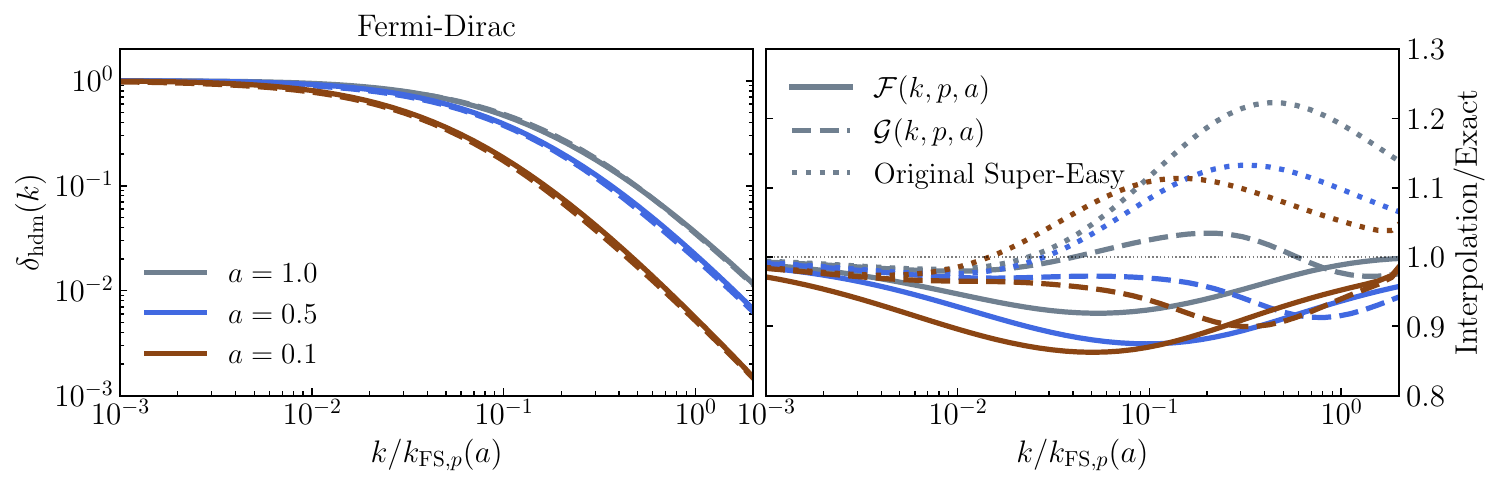}
\caption{{\it Left}: Exact solution of $\delta_{\rm hdm}$ (solid) from equation~\eqref{eq:df_sprime} and the SuperEasy
interpolation (dashed), where the HDM is taken to be a $m=0.31$~eV SM neutrino. We have assumed an external potential  $\Phi(s) = g(s) \Phi(s_{\rm i})$, where $g(s)$ is the $\Lambda$CDM growth factor for a reference cosmology with $\omega_{\rm m}=0.1424$ and $h=0.6766$.  The wave number $k$ is normalised to the $p$-dependent free-streaming scale $k_{{\rm FS},p}$ evaluated at $p = \sqrt{3/2} \, (\zeta(3)/\ln 2) \, T_{\nu,0}$, where $T_{\nu,0}=1.95$~K is the present-day SM neutrino temperature.
{\it Right}: $\delta_{\rm HDM}$ constructed from equation~\eqref{eq:dh1} using the ${\cal F}$ interpolation function~\eqref{eq:interpolationf} (solid) and from equation~\eqref{eq:dh2} using the ${\cal G}$ interpolation function~\eqref{eq:interpoliationG} (dashed), normalised to the exact solution at the same scale factors.
For completeness we also show the original integrated SuperEasy solution~\cite{Chen:2020kxi} (dotted).  Other choices of $m$ yield quantitatively similar results, but with the features shifted to the left (i.e., to lower $k$ values) for $m < 0.31$~eV and to the right for $m>0.31$~eV.}
\label{fig:interpolation-deltanu}
\end{figure}

\begin{figure}[t]
\centering
\includegraphics[width=\textwidth]{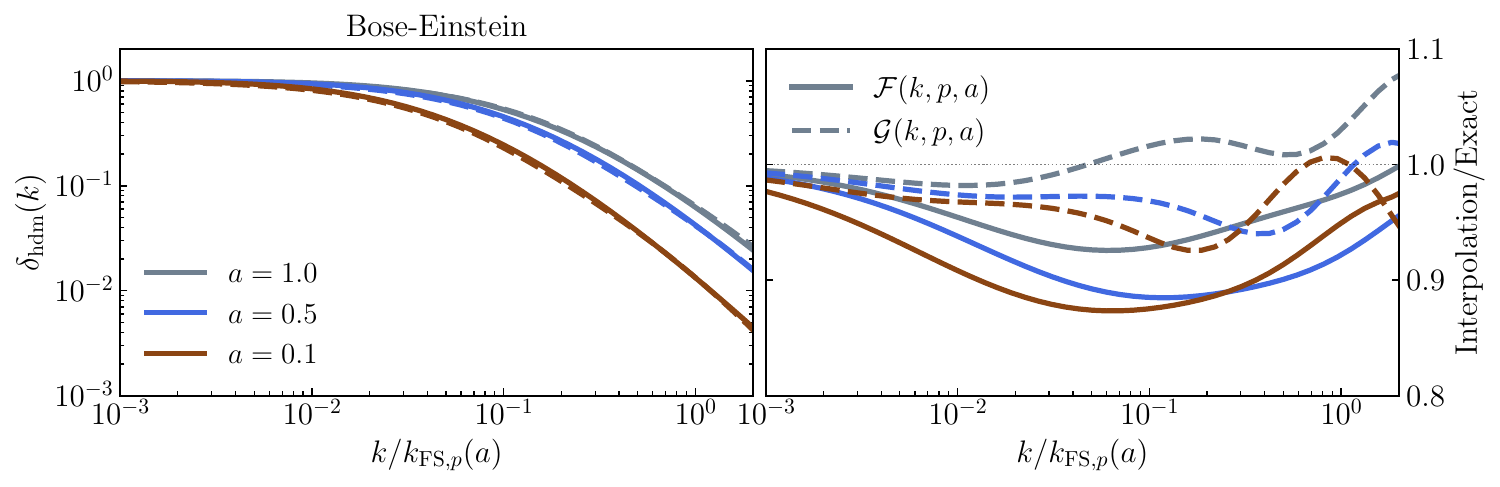}
\caption{Same as figure~\ref{fig:interpolation-deltanu}, but for a boson of mass $m=0.31$ following a relativistic Bose-Einstein distribution of present-day temperature $T_{\nu,0} = 1.95$~K.}
\label{fig:interpolation-deltaBE}
\end{figure}

Observe also that, given the limiting solutions of ${\cal F}(k,p,s)$ in equation~\eqref{eq:asymptotic}, the function~${\cal G}$ must always have the asymptotic behaviours 
\begin{equation}
{\cal G}(k/k_{{\rm FS},p}\to 0) \to 1,   \qquad
{\cal G}(k/k_{{\rm FS},p}\to \infty) \to   \frac{k_{{\rm FS},p}^2}{k^2},  
\label{eq:Gasyptotic}
\end{equation}
 independently of exactly how we interpolate the two limits of ${\cal F}(k,p,s)$.  Thus, at a practical level, we could equally interpolate ${\cal G}$ itself between these limits, without reference to ${\cal F}$ (as would be the case had we adopted the MFLR approach from the start). Indeed, figure~\ref{fig:interpolation-deltanu} shows that the interpolation function
\begin{equation}
\label{eq:interpoliationG}
   {\cal G}(k,p,s) =  \frac{k_{{\rm FS},p}^2}{k^2+ \beta  k k_{{\rm FS},p}+k_{{\rm FS},p}^2},
\end{equation} 
where $\beta=1$, in conjunction with equation~\eqref{eq:dh2} reproduces the exact HDM density contrast $\delta_{\rm hdm}$ for a $m=0.31$~eV SM neutrino to better than 10\% accuracy over all wave numbers $k$ of interest in the scale factor range $0.1\leq a \leq 1$.   At wave numbers below which free-streaming effects kick in, i.e., $k \lesssim 0.007\, h$/Mpc, the agreement is better than 3\%, outperforming the ${\cal F}$ interpolation construction~\eqref{eq:interpolationf} and~\eqref{eq:dh1}.  Similar comparisons can also be seen for a bosonic HDM of the same mass and temperature (figure~\ref{fig:interpolation-deltaBE}).
We shall therefore use the scheme~\eqref{eq:dh2} and its associated interpolation function~\eqref{eq:interpoliationG} in the rest of the paper to construct $\delta_{\rm hdm}$.

\subsection{Comparison with the original SuperEasy linear response for SM neutrinos}\label{sec:ose}

Before proceeding further with the generalised SuperEasy linear response approach~\eqref{eq:dh2}, let us also contrast the approach with the original SuperEasy linear response model~\cite{Ringwald:2004np,Wong:2008ws,Chen:2020kxi} for SM massive neutrinos. Relative to the {\it generalised} SuperEasy approach presented in this work, the {\it original} SuperEasy linear response interpolates approximate solutions of the momentum-integrated, total neutrino density contrast $\delta_\nu(\vec{k},s)$, in the clustering and the free-streaming limits.  That is, momentum-integration takes places {\it before} interpolation. From equations~\eqref{eq:dh2} and~\eqref{eq:Gasyptotic}, we find 
 the momentum-integrated limiting solutions 
\begin{equation}
\begin{aligned}
\label{eq:integrated}
\frac{\delta_\nu}{\delta_{\rm m}} (k/k_{\rm FS} \to 0,s) & \simeq 
\frac{1}{C} \int {\rm d} p\,p^2  \, \bar{f}(p) = 1, \\
\frac{\delta_\nu}{\delta_{\rm m}} (k/k_{\rm FS} \to \infty,s) & \simeq 
 \frac{1}{C} \int {\rm d} p\,p^2 \, \bar{f}(p) \frac{k_{{\rm FS},p}^2(s)}{k^2}=
\frac{k_{\rm FS}^2(s)}{k^2},
\end{aligned}
\end{equation}
where, for a relativistic Fermi-Dirac distribution $\bar{f}$, the integrated free-streaming scale $k_{\rm FS}$ has an exact analytical solution
\begin{equation}
k_{\rm FS}(s) \equiv \sqrt{ \frac{\ln 2}{\zeta(3)}}\, \frac{m_\nu}{T_{\nu}(s)} \, {\cal H}(s)\Omega^{1/2}_{\rm m}(s)  = \sqrt{ \frac{\ln 2}{\zeta(3)}}\, \frac{m_\nu}{T_{\nu,0}} \, H_0 \, \sqrt{a(s) \, \Omega_{{\rm m},0}}\, ,
\label{eq:intfs}
\end{equation}
with Riemann zeta function $\zeta(3)\simeq 1.202$, and $T_\nu$ is the neutrino temperature.  A simple function $k_{\rm FS}^2/(k + k_{\rm FS})^2$ interpolates between the two limits.

The right panel of figure~\ref{fig:interpolation-deltanu} shows this interpolation function next to the generalised SuperEasy models of this work.  Clearly, while the original SuperEasy  approach is also able to reproduce the exact $\delta_{\rm hdm}$ to better than 10\% at $k \lesssim 0.007\, h$/Mpc and better than 20\% across all wave numbers for $0.1 \leq a \leq 1$, overall the generalised SuperEasy interpolation function~\eqref{eq:interpoliationG} is superior.  The price we pay, of course, is an additional momentum-integration that must be performed numerically.   We stress however that the applicability of the original SuperEasy approach is limited: it works only in those cases where the integrated free-streaming scale $k_{\rm FS}$ can be defined and/or has an analytical solution.  For $\bar{f}(p)$ equal to the relativistic Bose-Einstein distribution, for example, $k_{\rm FS}$  as defined by the second integral in equation~\eqref{eq:integrated} is formally infinite.  For arbitrary forms of $\bar{f}(p)$, numerical methods are always necessary to evaluate $k_{\rm FS}$. 


\subsection{Generalised SuperEasy linear response to the cold matter density}

Returning now to the {\it generalised} SuperEasy approach, we note that
equation~\eqref{eq:dh2} represents the linear response of $\delta_{\rm hdm}(\vec{k},s)$ to the {\it total matter density contrast} $\delta_{\rm m}(\vec{k},s)$.  The next step is to recast it as a linear response to the {\it cold matter density contrast} $\delta_{\rm cb}(\vec{k},s)$, since it is normally $\delta_{\rm cb}(\vec{k},s)$ that is readily calculable (using, e.g., $N$-body simulations or higher-order perturbation theory).  To do so, we discretise the integral~\eqref{eq:dh2} and rewrite it as a sum over $N$ momentum bins in the following manner:
\begin{equation}
\label{eq:discretise}
\int_0^\infty {\rm d} p \, p^2  \bar{f}(p)\, {\cal G}(k,p,s) \to \sum_{i=1}^N \left[\int_0^\infty {\rm d} p \, p^2 \bar{f}(p) \, \omega_i(p) \right] {\cal G}_i(k,s),
\end{equation}
where $\omega_i(p)$ is the weight of the $i$-th momentum-bin, and ${\cal G}_i(k,s) \equiv {\cal G}(k,p_i,s)$ is the interpolation function of that bin whose free-streaming scale $k_{{\rm FS},i}$ is to be evaluated at a representative momentum $p_i$.   The weight function $\omega_i(p)$ is user-defined and can be as simple as a top hat $\omega_i(p) = \Theta(p-p_{{\rm min},i}) \Theta(p_{{\rm max},i}-p)$, with $\Theta(x)$ the Heaviside function.

Discretising as per equation~\eqref{eq:discretise}  is akin to having $N$ different types of HDM, each defined by its representative free-streaming scale $k_{{\rm FS}_i}$ and has a density contrast $\delta_{{\rm h}_i}$.  Then, defining the $i$-th HDM fraction as
\begin{equation}
f_{{\rm h}_i} \equiv  \frac{f_{\rm hdm}}{C }\int_0^\infty {\rm d} p \, p^2 \bar{f}(p)\, \omega_i(p),
\label{eq:fhi}
\end{equation}
which satisfies the normalisation condition $f_{\rm hdm} \delta_{\rm hdm} = \sum_{i=1}^N f_{{\rm h}_i} \delta_{{\rm h}_i}$, the density contrast of the $i$-th HDM is now formally given by
\begin{equation}
    \delta_{{\rm h}_i}(\vec{k},s)=\mathcal{G}_i(k,s)\delta_{\rm m}(\vec{k},s)  =\mathcal{G}_i(k,s)\left[f_{\rm cb}\delta_{\rm cb}(\vec{k},s)+\sum_{j=1}^Nf_{{\rm h}_j}\delta_{{\rm h}_j}(\vec{k},s)\right]
    \label{eq:dh3}
\end{equation}
in the linear response solution. 

Next, we isolate the $i$-th density contrast $\delta_{{\rm h}_i}$ in equation~\eqref{eq:dh3}  by writing $\sum_{j=1}^Nf_{{\rm h}_j}\delta_{{\rm h}_j}= f_{{\rm h}_i}\delta_{{\rm h}_i}  + \sum_{j\neq i}^Nf_{{\rm h}_j}\delta_{{\rm h}_j}$.  Rearranging~\eqref{eq:dh3} in favour of  $\delta_{{\rm h}_i}$ then yields 
\begin{equation}
\label{eq:deltaai2}
    \delta_{{\rm h}_i}(\vec{k},s)=L_i(k,s) \left[f_{\rm cb} \delta_{\rm cb}(\vec{k},s)+\sum_{j\neq i}^Nf_{{\rm h}_j}\delta_{{\rm h}_j}(\vec{k},s)\right],
\end{equation} 
where $L_i(k,s)\equiv \mathcal{G}_i(k,s)/[1-\mathcal{G}_i(k,s)f_{{\rm h}_i}]$.
Following~\cite{Chen:2020kxi}, we define $\vec{\delta_{\rm h}} \equiv \left(\delta_{{\rm h}_1},\delta_{{\rm h}_2},\cdots,\delta_{{\rm h}_N} \right)^T$,
$\vec{L} \equiv \left(L_1,L_2,\cdots,L_N \right)^T$, and 
\begin{equation}
 \hat{M} \equiv\begin{pmatrix}
        0 & L_1f_{{\rm h}_2} & L_1f_{{\rm h}_3} & \dots & L_1f_{{\rm h}_N}  \\
        L_2f_{{\rm h}_1} & 0 & L_2f_{{\rm h}_3} & \dots & L_2f_{{\rm h}_N}\\
        L_3f_{{\rm h}_1} & L_3f_{{\rm h}_2} & 0 & \dots & L_3f_{{\rm h}_N}\\
        \vdots & \vdots & \vdots & \ddots & \vdots\\
        L_Nf_{{\rm h}_1} & L_Nf_{{\rm h}_2} & L_Nf_{{\rm h}_3} & \dots & 0\\
    \end{pmatrix}.
    \label{eq:mmatrix}
\end{equation}
Then, equation~\eqref{eq:deltaai2} can be equivalently expressed in matrix notation 
\begin{equation}
    \vec{\delta}_{\rm h}(\vec{k},s)=\left[\hat{I}_N-\hat{M}(k,s)\right]^{-1}\vec{L}(k,s)f_{\rm cb}\delta_{\rm cb}(\vec{k},s),
\end{equation}
where $\hat{I}_N$ is the $N \times N$ identity matrix,
and the total HDM density contrast $ \delta_{\rm hdm}$ can be constructed from the normalisation condition $f_{\rm hdm} \delta_{\rm hdm} = \sum_{i=1}^N f_{{\rm h}_i} \delta_{{\rm h}_i}$ to give
\begin{equation}
f_{\rm hdm} \delta_{\rm hdm} (\vec{k},s) = {\vec{f}_{\rm h}}^{~T}  \left[\hat{I}_N-\hat{M}(k,s)\right]^{-1}\vec{L}(k,s)f_{\rm cb}\delta_{\rm cb} (\vec{k},s),
\label{eq:totaldeltaa}
\end{equation}
with $\vec{f}_{\rm h} \equiv \left(f_{{\rm h}_1},f_{{\rm h}_2},\cdots,f_{{\rm h}_N} \right)^T$.  Equation~\eqref{eq:totaldeltaa} is thus the generalised SuperEasy linear response  of the  total HDM density contrast $\delta_{\rm hdm}(\vec{k},s)$ to a dominant cold matter density contrast~$\delta_{\rm cb}(\vec{k},s)$ at the same wave vector~$\vec{k}$ and time~$s$.


\subsection{Generalised SuperEasy gravitational potential}

Substituting the linear response solution~\eqref{eq:totaldeltaa} for $\delta_{\rm hdm}$ into the Poisson equation~\eqref{eq:poisson}, we can immediately write down the $k$-space Poisson equation in the linear response regime,
\begin{equation}
k^2 \Phi(\vec{k},s)\simeq -\frac{3}{2} {\cal H}^2 \, \Omega_{\rm cb}(s) \left\{1+ {\vec{f}_{\rm h}}^{~T}  \left[\hat{I}_N-\hat{M}(k,s)\right]^{-1} \vec{L}(k,s) \right\} \delta_{\rm cb} (\vec{k},s),
\label{eq:poisson1}
\end{equation}
which features $\delta_{\rm cb}(\vec{k},s)$ as the only real-time variable, and is correct to all order in $f_{{\rm h}_i}$.  Note that it is the background cold matter density $\Omega_{\rm cb}(s)$ that appears in the prefactor, as would be the case in, e.g., an $N$-body simulation code. 

Formally, 
equation~\eqref{eq:poisson1} is a closed-form expression that can be immediately evaluated upon specifying a value for $\delta_{\rm cb}(\vec{k},s)$ (e.g., from an $N$-body simulation).  In practice, however, its evaluation
involves a matrix inversion that needs in principle to be performed at every time step and at every wave number $k$ of interest.  This can be resource-intensive, particularly if the number of momentum bins $N$ is large.  Fortunately, however, all non-zero entries in the matrix $\hat{M}$ are at most ${\cal O}(f_{{\rm h}_i})$ and as a subdominant HDM component, $f_{{\rm hdm}} \ll 1$ is guaranteed.  We can therefore expand around the small parameters $f_{{\rm h}_i}$ to obtain 
\begin{equation}
\left[\hat{I}_N-\hat{M}(k,s) \right]^{-1} \simeq \hat{I}_N+\hat{M}(k,s) + \hat{M}(k,s) \hat{M}(k,s) + {\cal O}(f_{{\rm h}_i}^3),
\label{eq:noinversion}
\end{equation}
and thereby circumvent the need for a matrix inversion operation at every $k$ and $s$.

\begin{figure}[t]
    \centering
    \includegraphics[width=0.9\textwidth]{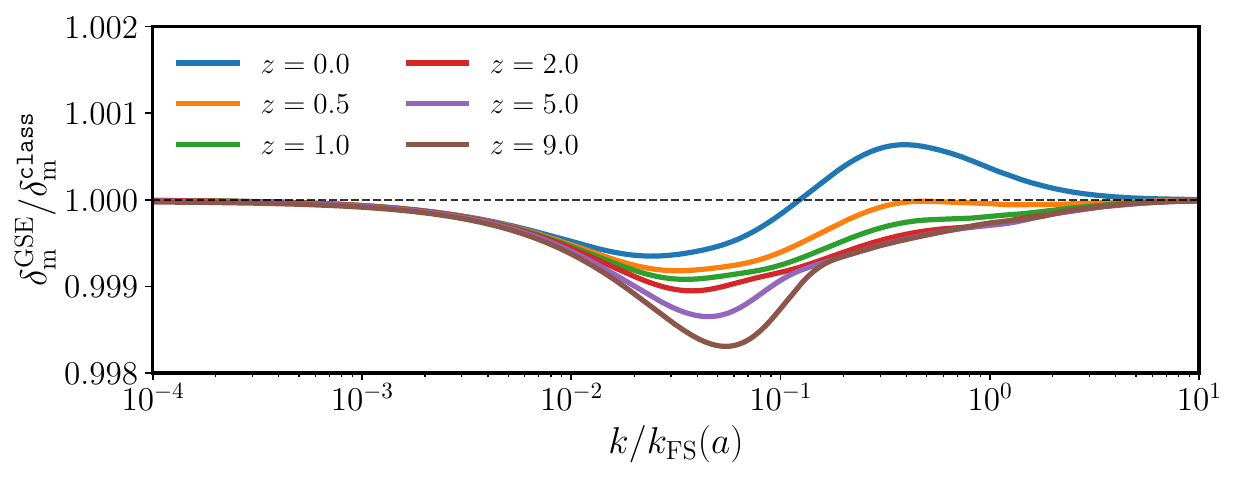}
    \caption[]{Total matter density contrast $\delta^{\rm GSE}_{\rm m}$ at $0 \leq z \leq 9$ computed using the generalised SuperEasy method (equation~\eqref{eq:gse_poisson}) from a given $\delta_{\rm cb}$,
    compared to the prediction of  \codo{class} $\delta^{\codo{class}}_{\rm m}=f_{\rm cb}\delta^{\codo{class}}_{\rm cb}+f_{\rm hdm}\delta_{\rm hdm}^{\codo{ class}}$ for the same $\delta_{\rm cb} = \delta^{\codo{class}}_{\rm cb}$.
The HDM is taken to be three degenerate SM neutrinos with $\sum m_{\nu}=0.465$~eV.  The 
    \codo{class} run has been performed at high accuracy and without fluid approximations, while the generalised SuperEasy computation uses $N=15$ Gauss-Laguerre momentum bins.
   The wave number $k$ is normalised  to the integrated free-streaming scale $k_{\rm FS}(a)$ defined in equation~\eqref{eq:intfs}.}
    \label{fig:gseclass}
\end{figure}

Then, to arrive at a simplified gravitational potential we need simply to substitute the approximation~\eqref{eq:noinversion} into the Poisson equation~\eqref{eq:poisson1}.  The details of the calculation can be found in appendix~\ref{app:derivations}.  Here, we present the leading-order result in $f_{{\rm h}_i}$:
\begin{equation}
\begin{aligned}
k^2 \Phi(\vec{k},s)  & \simeq -\frac{3}{2} {\cal H}^2 \,  \Omega_{\rm m}(s) \left\{
1+  \sum_{i=1}^N f_{{\rm h}_i} \big[{\cal G}_i(k,s) -1\big]  + {\cal O}(f_{{\rm h}_i}^2) \right\} \delta_{\rm cb} (\vec{k},s) \\
& \equiv -\frac{3}{2} {\cal H}^2 \, \Omega_{\rm cb}(s)\,  \tilde{g}(k,s) \, \delta_{\rm cb} (\vec{k},s) .
\label{eq:gse_poisson}
\end{aligned}
\end{equation} 
Equation~\eqref{eq:gse_poisson} is the centrepiece of the generalised SuperEasy linear response approach, where $\Omega_{\rm m} = \Omega_{\rm cb}/f_{\rm cb}$, and we observe that
the modification factor $\{\cdots\}=  f_{\rm cb}\, \tilde{g}(k,s)$---corresponding to the SuperEasy prediction of $\delta_{\rm m}/\delta_{\rm cb}$---modulates from unity as $k \to 0$ to $1-\sum_{i=1}^N f_{{\rm h}_i} = 1-f_{\rm hdm}$ as $k \to \infty$ as required.  As already discussed in section~\ref{sec:ose}, relative to the original SuperEasy linear response, the generalised SuperEasy approach incurs an extra summation over the HDM momentum.  However, with a judicious choice of weights $\omega_i(p)$ and hence HDM fractions $f_{{\rm h}_i}$, the number of momentum bins required to capture the modification factor to sub-0.2\% accuracy can be kept to no more than 10 to 20, as can be seen in figure~\ref{fig:gseclass} where we compare $f_{\rm cb} \, \tilde{g}(k,s)$ with the prediction of $\delta_{\rm m}/\delta_{\rm cb}$ from \codo{class} at $0.1 \leq a \leq 1$, using 15 momentum bins in the Gauss-Laguerre quadrature scheme (see section~\ref{sec:LRsims} for details).

Finally, while we have presented the generalised SuperEasy linear response in these sections as though there was only one HDM species (albeit with a non-trivial phase space distribution), extending the method to include multiple HDM species of different masses, temperatures, and/or statistics in a ``mixed HDM'' setting is straightforward. At the most basic level one could simply add any extra species to the sum in the generalised SuperEasy gravitational potential~\eqref{eq:gse_poisson} with the appropriate weight $f_{{\rm h}_i}$.  We note however that it is also possible to exploit overlapping kinematic properties of the different HDM species to combine them into one single effective HDM species for a more efficient numerical evaluation; see section~\ref{sec:n-body_simulations} and the companion paper~\cite{Flows2} for details.


\section{Application to mixed HDM in \texorpdfstring{$N$}{N}-body simulations}
\label{sec:n-body_simulations}

\begin{table}[t]
    \centering
    \begin{tabular}{c|cccccc}
        \midrule
        \midrule
         Model & Content  & $\sum m_{\nu}$/eV & $m_{\rm hdm}$/eV & $T_{{\rm hdm},0}$/K & $\Omega_{{\rm hdm},0}$ & $\Delta N_{\rm eff}$ \\
         \midrule
        E1 & SM neutrinos\Plus{}axions  & 0.059 & 0.048 & 1.88 & 0.0025 & 0.19 \\
        E2 & SM neutrinos           & 0.161 & 0.054 & 1.95 & 0.0038 & / \\
        \midrule
        E3 & SM neutrinos\Plus{}BE     & 0.059 & 0.074 & 1.88 & 0.0045 & 0.48 \\
        E4 & SM neutrinos          & 0.315 & 0.105 & 1.95 & 0.0074 & / \\
        \midrule
        \midrule
    \end{tabular}
    \caption{The four mixed HDM models considered in this work and their respective energy content. All four models contain three normally-ordered SM neutrinos with total mass $\sum m_\nu$. Model E1 contains in addition a QCD axion of mass $m_a=0.23$~eV and temperature~$T_{a,0} = 1.86$~K following the calculation of~\cite{Notari:2022ffe}, while E3 has a generic thermal boson of the same mass and temperature. These models predict a radiation excess at early times specified by a non-zero $\Delta N_{\rm eff}$. 
    The neutrino-only models E2 and E4 have been chosen by matching their the small-scale linear matter power spectra to those of E1 and E3, such that $\sigma_8=0.75$ for E1/E2 and $\sigma_8=0.79$ for E3/E4.  Within the effective HDM description (see section~\ref{sec:effectiveHDM}), these models are specified by the effective HDM mass $m_{\rm hdm}$, effective HDM temperature~$T_{\rm hdm}$, and the present-day reduced total HDM energy density $\Omega_{{\rm hdm},0}$.}
    \label{tab:cosmologies}
\end{table}

We demonstrate in this section how to use the generalised SuperEasy method in $N$-body simulations.
Since the method is entirely agnostic of the HDM phase space distribution, we apply it to simulate several mixed HDM cosmologies.  In particular, we consider cases wherein the HDM component always comprises three normally-ordered SM neutrinos and may include in addition a thermally-produced QCD axion as computed in reference~\cite{Notari:2022ffe}%
\footnote{In order to generate a thermal relic abundance of QCD axions substantial enough to be observed as a hot dark matter, the axion mass needs to exceed meV values. In such cases, the QCD axion cannot simultaneously explain the cold dark matter content of the universe.} 
or a generic thermal boson.  The specifics of the HDM sectors of these cosmologies are given in table~\ref{tab:cosmologies}, to be discussed in more detail below in section~\ref{sec:effectiveHDM}.  As usual we hold all  non-HDM parameters and particularly the total matter density $\Omega_{\rm m}$ fixed across cosmologies, i.e., an increase in $\Omega_{\rm hdm}$ is compensated for by a decrease in $\Omega_{\rm cb}$, to ensure all models have matching power spectra on large scales.  Specifically, we used the Planck 2018 values~\cite{Planck:2018vyg}
\begin{equation}
    \Omega^0_{\rm m}=0.311, ~~~\Omega^0_{\rm b}=0.049, ~~~A_s=2.1\times 10^{-9}, ~~~n_s=0.966, ~~~h=0.6766, 
    \label{eq:planckparams}
\end{equation} 
for the baryon density $\Omega_{\rm b}$, the primordial fluctuation amplitude $A_s$ and spectral index $n_s$, and the reduced Hubble parameter~$h$.

In the following we describe first in section~\ref{sec:effectiveHDM} how we combine the mixed HDM into a single effective HDM, before detailing our simulations and results in sections~\ref{sec:LRsims}--\ref{sec:hybrid}.


\subsection{Effective HDM}
\label{sec:effectiveHDM}

Conventionally, tracking multiple HDM species with different particle masses requires that we solve multiple Vlasov equations, each with a specific particle mass and homogeneous background phase space distribution.  In the non-relativistic Newtonian framework, the total HDM energy density at any time is then given by
\begin{equation}
\begin{aligned}
    \rho_{\rm hdm}(\vec{x},s) & = \frac{1}{(2\pi a)^{3}}\sum_{\alpha=1}^{N_\alpha} m_\alpha \, g_\alpha \int_0^{\infty} {\rm d}^3p \, 
      f_\alpha^{(p)} (\vec{x},\vec{p},s) \\
        & = \frac{1}{(2\pi a)^{3}}\sum_{\alpha=1}^{N_\alpha} m_\alpha \, g_\alpha\,  T_{\alpha,0}^3\int_0^{\infty} {\rm d}^3q_\alpha \, 
      f_\alpha^{(q)} (\vec{x},\vec{q}_\alpha,s),
      \end{aligned}
\end{equation}
where $T_{\alpha,0}$ is the present-day temperature of the $\alpha$-th HDM species, $g_\alpha$ its internal degrees of freedom, and we have defined an $\alpha$-dependent dimensionless momentum $\vec{q}_\alpha \equiv \vec{p}/T_{\alpha,0}$ in the second line, a quantity commonly used in linear Boltzmann codes like \codo{camb} and \codo{class}.
The superscripts $(p)$ and $(q)$ have been inserted to highlight that $f_\alpha^{(p)}$ and $f_\alpha^{(q)}$ have different functional forms.

However, as discussed in the companion paper~\cite{Flows2}, because free-streaming is a kinematic effect dependent only on the HDM particle {\it velocity} rather than its {\it momentum}, it is convenient to express the HDM phase space densities in terms of the Lagrangian velocity  $\vec{v}\equiv \vec{p}/m_\alpha$ in lieu of the comoving momentum~$\vec{p}$, i.e., $f_\alpha^{(p)}(\vec{x},\vec{p},s) \to f_\alpha^{(v)}(\vec{x},\vec{v},s)$.
Then, with $\{\vec{x},\vec{v},s\}$ as independent variables, all dependencies
on the particle mass $m_\alpha$ in the Vlasov equation~\eqref{Eq:BoltzmannEq} must immediately drop out, and multiple (non-relativistic) HDM phase space distributions can be equivalently combined via 
\begin{equation}
    \rho_{\rm hdm}(\vec{x},s)  = \frac{1}{(2\pi a)^{3}} \int_0^{\infty} {\rm d}^3v \, 
    \sum_{\alpha=1}^{N_\alpha} m^4_\alpha \, g_\alpha\,   f_\alpha^{(v)} (\vec{x},\vec{v},s) \equiv  \frac{1}{(2\pi a)^{3}} \int_0^{\infty} {\rm d}^3v \, F^{(v)}(\vec{x},\vec{v},s), 
    \label{eq:combinedv}
\end{equation}
where the grand distribution $F^{(v)}(\vec{x},\vec{v},s)$ can be tracked using a single Vlasov equation.%
\footnote{These statements remain generally true also in the fully relativistic case, with the important feature that the correct velocity variable is still the Lagrangian velocity $v=p/m_\alpha$ and {\it not} the physical particle speed $p/\sqrt{p^2+a^2 m_\alpha^2}$~\cite{Flows2}.}

In practical applications, however, rather than integrating in $v$ it is useful to retain an $\alpha$-{\it in}dependent dimensionless momentum $q \equiv (m_{\rm hdm}/T_{{\rm hdm},0}) \, v = (m_\alpha/T_{\alpha,0}) \, v $ as the integration variable (because this is how linear Boltzmann codes are written), normalised to an effective HDM temperature $T_{{\rm hdm},0}$ and an effective HDM mass $m_{\rm hdm}$, such that equation~\eqref{eq:combinedv} can also be written as
\begin{equation}
\begin{aligned}
    \rho_{\rm hdm}(\vec{x},s) &  = \frac{m_{\rm hdm} T^3_{{\rm hdm},0}}{(2\pi a)^{3}} \int_0^{\infty} {\rm d}^3q \, 
    \sum_{\alpha=1}^{N_\alpha} \left(\frac{m_\alpha}{m_{\rm hdm}}\right)^4  g_\alpha\,   f_\alpha^{(q)} \left(\vec{x},\frac{m_\alpha}{m_{\rm hdm}} \frac{T_{{\rm hdm},0}}{T_{\alpha,0}} \vec{q},s \right) \\
    &\equiv  \frac{m_{\rm hdm} T^3_{{\rm hdm},0}}{(2\pi a)^{3}} \int_0^{\infty} {\rm d}^3q \, F^{(q)}(\vec{x},\vec{q},s).
    \label{eq:combinedv-rewr}
    \end{aligned}
\end{equation}
The choices of $T_{{\rm hdm},0}$ and $m_{\rm hdm}$ are arbitrary.  We find it convenient to use an effective HDM mass defined via the homogeneous background phase space distributions~$\bar{f}_\alpha^{(q)}$,
\begin{equation}
m_{\rm hdm} \equiv \frac{\bar{\rho}_{\rm hdm}}{\bar{n}_{\rm hdm}}=\frac{\sum_{\alpha=1}^{N_\alpha} m_\alpha \, g_\alpha \, T_{\alpha,0}^3 \int_0^{\infty} {\rm d}q_\alpha \, q_\alpha^2\, 
      \bar{f}_\alpha^{(q)} (q_\alpha) }{\sum_{\alpha=1}^{N_\alpha}  g_\alpha \, T_{\alpha,0}^3\int_0^{\infty} {\rm d} q_\alpha \,q_\alpha^2 \, 
      \bar{f}_\alpha^{(q)}  (q_\alpha) },
\end{equation}
and a mass-weighted average effective HDM temperature 
\begin{equation}
    m_{\rm hdm}\, T_{{\rm hdm},0}=\frac{1}{N_\alpha} \sum_{\alpha=1}^{N_\alpha} m_\alpha \, T_{\alpha,0}.
\end{equation} 
We refer to reference~\cite{Flows2} for further details on the effective HDM treatment.

\begin{figure}[t]
    \centering
    \includegraphics[width=0.6\textwidth]{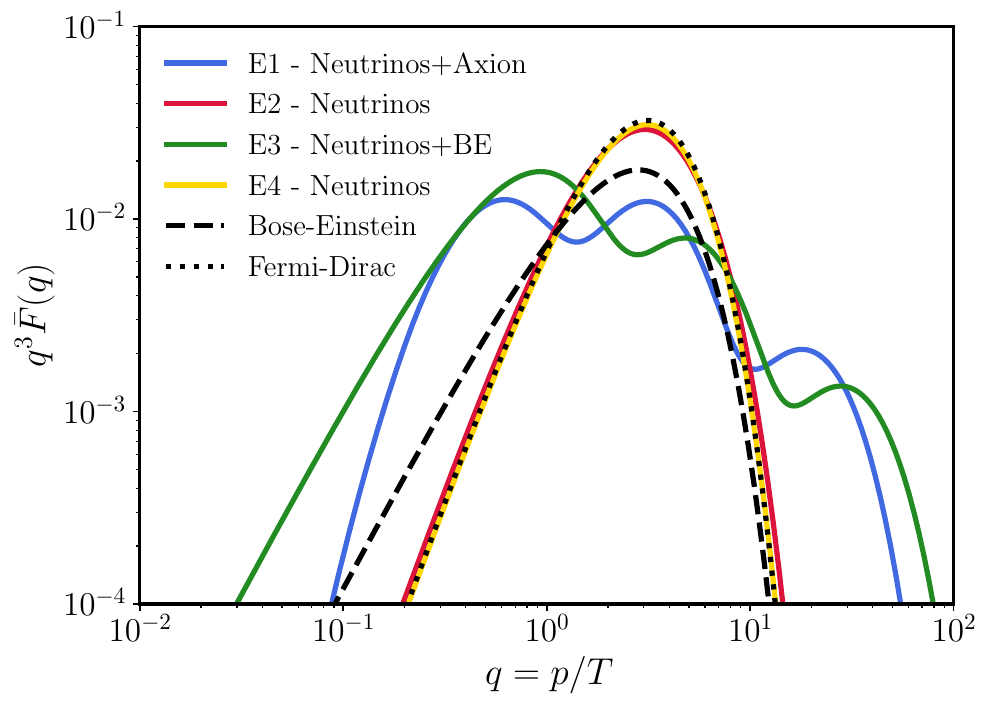}
    \caption{The effective homogeneous phase space distributions $\bar{F}(q)$ of the four HDM models (solid lines) summarised in table~\ref{tab:cosmologies}.  For reference we also show alongside a relativistic Fermi-Dirac (dotted) and a relativistic Bose-Einstein (dashed) distribution.     
    As explained in the text, while the low-momentum tail of E3 (green; neutrinos\Plus{}BE) is consistent with a pure Bose-Einstein distribution, the E1 model (blue; neutrinos\Plus{}axion~\cite{Notari:2022ffe}) is distorted in the same regime and in fact more closely resembles the Fermi-Dirac form.  On the other hand, the two neutrino-only models E2 and E4 (red and yellow) have distributions highly consistent with the Fermi-Dirac form
    (because of their relatively large $\sum m_\nu$ values, such that the neutrino mass spectra approach the degenerate limit).
    }
    \label{fig:psd}
\end{figure}

Applied to our mixed HDM cosmologies listed in table~\ref{tab:cosmologies}, we first of all remind the reader that our models always contain three SM neutrinos with normally-ordered masses constrained by neutrino oscillation experiments.  These neutrinos always have the same temperature $T_{\nu, 0}\simeq 1.95~{\rm K}$.  Details of the individual model and their motivation are as follows.
\begin{itemize}
    \item \textbf{E1:}  This model contains normally-ordered SM neutrino held at their minimum allowed masses, $m_{3,2,1}=50,9,0$~meV,%
\footnote{These mass values follow from the squared mass splittings obtained from three-flavour global fits of recent neutrino oscillation data: $\Delta m_{21}^2 \equiv m_2^2-m_1^2 \simeq (7.5^{+0.22}_{-0.20}) \times 10^{-5}~{\rm eV}^2$ and $\Delta m_{31}^2 \equiv m_3^2-m_1^2 \simeq (2.5^{+0.02}_{-0.03}) \times 10^{-3}~{\rm eV}^2$~\cite{deSalas:2020pgw,Esteban:2020cvm,Capozzi:2017ipn}.}
 plus a QCD axion of mass $m_a =0.23$~eV (corresponding to an axion decay constant $f_a=2.5\times 10^7$~GeV).
The production of these axions in the early universe has been computed in reference~\cite{Notari:2022ffe} using the momentum-dependent collisional Boltzmann equation and incorporating model-independent thermal production rates derived from phenomenological data. 

Because the universe's entropy $g_{*s}$ varies greatly across the axion production epoch and high-momentum axions decouple later than low-momentum ones, the authors of~\cite{Notari:2022ffe} find significant deviations in the homogeneous axion phase space distribution from a pure Bose-Einstein distribution (see figure~\ref{fig:psd}).  They further find an upper limit of $m_a \lesssim 0.24$~eV~\cite{Notari:2022ffe} from cosmological data; our choice of $m_a =0.23$~eV sits just within the allowed range.  Then, together with an axion temperature of $T_{a,0} = 1.86$~K, we determine the scenario's effective HDM temperature and effective HDM mass to be $T_{{\rm hdm},0} = 1.88$~K and 
$m_{\rm hdm}=48$~meV respectively.%
\footnote{We thank the authors of reference~\cite{Notari:2022ffe} for providing us with their numerical evaluations of the $f_a=2.5\times 10^7$~GeV axion momentum distribution.}

\item \textbf{E2:} This model comprises only normally-ordered SM neutrinos totalling $\sum m_{\nu}=0.161$~eV, where the masses have been tuned (while respecting oscillation constraints) such that at $z=0$ the model's {\it linear} total matter power spectrum on small scales matches that of E1.  The effective HDM mass of this scenario is $m_{\rm hdm}=54$ meV. 

Enforcing the same small-scale suppression across different HDM cosmologies generally leads to a mismatch in the present-day total HDM density $\Omega_{{\rm hdm},0}$, which indicates, unsurprisingly, that the parameter $\Omega_{{\rm hdm},0}$ alone does not fully capture the impact of HDM even at the linear level.  Additionally, matching the linear suppression as we have done here has the benefit of isolating the effects of non-linear evolution.

\item \textbf{E3:} Given the axion distribution's departure from pure Bose-Einstein statistics---as shown in figure~\ref{fig:psd}, the slope in the low-$p$ regime is in fact closer to the relativistic Fermi-Dirac form---we consider also the case of a generic thermal boson (referred to as ``boson'' in the following) of the same mass and temperature as in E1 ($m_{\rm BE}=0.23$~eV, $T_{{\rm BE},0}=1.86$~K), whose homogeneous phase space density follows the relativistic Bose-Einstein distribution.
The model has in addition three normally-ordered SM neutrinos with the minimum mass sum $\sum m_{\nu}=59$ meV, giving an effective HDM mass and temperature of $m_{\rm hdm}=74$ meV and $T_{{\rm hdm},0} = 1.88$~K.

Note that this model is in fact already disfavoured by cosmological data~\cite{Notari:2022ffe}: the boson population leads to an excess of relativistic energy at CMB times, or, equivalently, $\Delta N_{\rm eff} = 0.48$ in terms of change in the effective number of neutrinos.  However, because the Bose-Einstein distribution has a stronger contribution from the small-velocity tail (see figure~\ref{fig:psd}), the model will have stronger non-linearities in the HDM sector.  This makes for an interesting validity test of linear response methods.

\item \textbf{E4:} The counterpart of E3, this model comprises three normally-ordered SM neutrinos of mass sum $\sum m_{\nu}=0.315$~eV tuned to match the small-scale linear total matter power spectrum of E3 at $z=0$. The effective HDM mass of the model is $m_{\rm hdm}=0.105$~eV. 
\end{itemize} 
Figure~\ref{fig:psd} shows the combined homogeneous background phase space distributions of these four HDM models.

\subsection{Generalised SuperEasy linear response simulations}\label{sec:LRsims}

As with the original SuperEasy approach~\cite{Chen:2020kxi}, the generalised SuperEasy linear response method of this work can be easily incorporated into any cosmological $N$-body simulation code that employs a PM gravity solver, by replacing on the Fourier grid the standard $k$-space Poisson equation with equation~\eqref{eq:gse_poisson} and identifying $\delta_{\rm cb}(\vec{k},s)$ with the $k$-space cold matter density contrast smoothed onto the PM-grid points. We have carried out this implementation in the TreePM code~\codo{gadget-4}~\cite{Springel:2020plp}.%
\footnote{Our modified version of \codo{gadget-4} is publicly available at {\tt \href{https://github.com/cppccosmo/gadget-4-cppc}{github.com/cppccosmo/gadget-4-cppc}}~.}

\begin{figure}[t]
    \centering
     \includegraphics[width=0.6\textwidth]{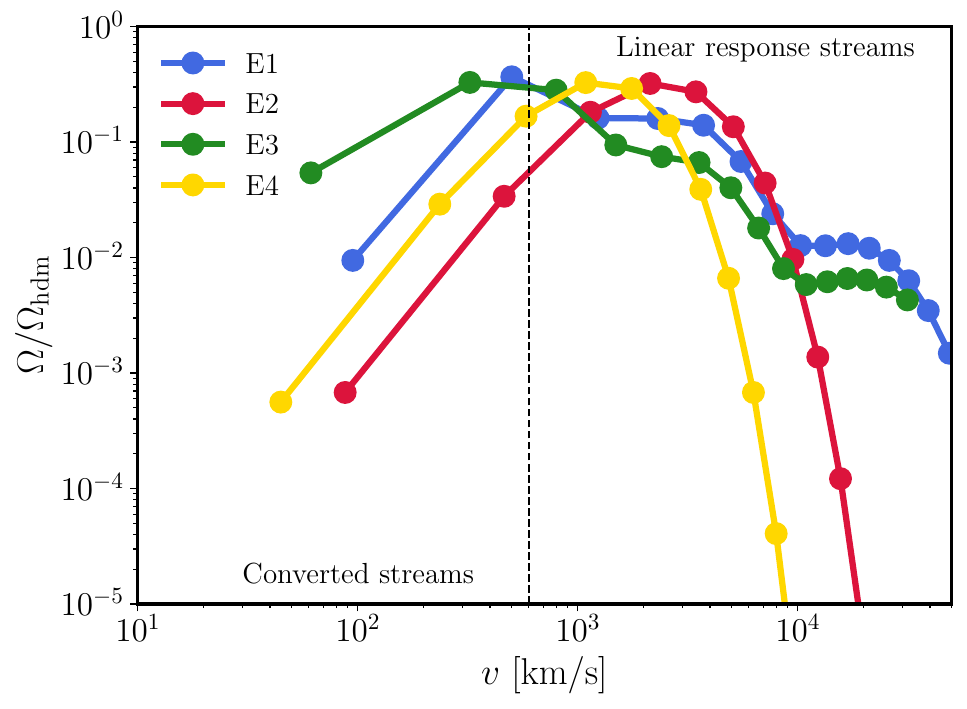}
    \caption{Partitioning of the effective HDM of the four models in table~\ref{tab:cosmologies} into $N_i=15$ momentum bins in the Gauss-Laguerre scheme.  The binning applies both to the generalised SuperEasy simulations of section~\ref{sec:LRsims} and to the flows of the MFLR/Hybrid simulations discussed in sections~\ref{sec:compareLRsims} and~\ref{sec:hybrid}.
We plot the HDM fraction $f_{{\rm h}_i}= \Omega_i/\Omega_{\rm hdm}$ in each bin/flow in terms of the present-day bin velocity $v = p/m_{\rm hdm} = q\, (T_{\rm hdm}/m_{\rm hdm})$ instead of the bin's dimensionless momentum $q$, because the conversion of MFLR flows to particles in the Hybrid approach is based on velocity and applies to flows with $v< 600$~km/s as established in~\cite{Chen:2022dsv}.
This conversion criterion is indicated by the vertical dotted line.}
    \label{fig:psd2}
\end{figure}

Our HDM momentum bin weights $\omega_i$ (see equations~\eqref{eq:discretise} and~\eqref{eq:fhi}) are set following the Gauss-Laguerre quadrature scheme, in which
the homogeneous background number density integral is approximated as 
\begin{equation}
    \int_0^{\infty} {\rm d}q\, q^2 \, \bar{F}(q)\simeq \sum_{i=1}^{n} W_i \, q_i^2\, e^{q_i}\, \bar{F}(q_i), 
    \label{eq:GLQ_rho_approx}
\end{equation} 
where $n$ is the number of Gauss-Laguerre points, $q_i$ the $i$-th root of the $n$-th Laguerre polynomial ${\cal L}_n(q)$, $\bar{F}$ the effective HDM distribution,  and the weights are given by
\begin{equation}
W_i = \frac{q_i}{(n+1)^2 [{\cal L}_{n+1}(q_i)]^2}.
\end{equation} In general, the number of Gauss-Laguerre points $n$ and the number of momentum bins used~$N_i$ need not coincide, as long as $n\geq N_i$.  However, we find that $N_i=n=15$ suffices to approximate the background number density integral to sub-0.01\% for models E2 and E4;
for models E1 and E3, the corresponding errors are within $0.6\%$ and $0.3\%$, respectively.  We therefore use these settings throughout this work.   In this binning scheme the HDM fraction $f_{{\rm h}_i}$  in the Poisson equation~\eqref{eq:gse_poisson} is given by
\begin{equation}
f_{{\rm h}_i} = f_{\rm hdm}\, \frac{W_i\, q_i^2\, e^{q_i}\bar{F}(q_i)}{\sum_j  W_j\, q_j^2\, e^{q_j} \, \bar{F}(q_j)}.
\end{equation} 
 Figure~\ref{fig:psd2} shows the momentum bins of the four HDM models of table~\ref{tab:cosmologies}, expressed in terms of the present-day bin velocity $v_i = p_i/m_{\rm hdm} = q_i\, (T_{\rm hdm}/m_{\rm hdm})$---for this reason the bins do not appear to coincide across cosmologies.

We perform simulations using $N_{\rm cb}=512^3$ particles to represent the cold matter, in a box of size length $L=256$ Mpc$/h$ and a PM grid of $N_{\rm PM}=1024^3$ cells. The cold particles are initialised at $z=99$ using the Zel'dovich approximation and the same back-scaling method as in~\cite{Chen:2020kxi}, where the $z=0$ linear matter spectrum is computed using \codo{class}. The softening length is set in units of the mean particle separation, namely, 
\begin{equation}
    \ell_{\rm soft}=C_{\rm soft} \frac{L}{N_{\rm PM}^{1/3}}, 
\end{equation} 
with $C_{\rm soft}=0.04$, leading to $\ell_{\rm soft}=10$ kpc$/h$ for $L=256$ Mpc$/h$. The cold matter power spectrum $P_{\rm cb}(k)$ is a standard output of \codo{gadget-4}; To construct the total matter power spectrum $P_{\rm m}(k)$ within the generalised SuperEasy approach, we simply multiply $P_{\rm cb}(k)$ by the modification factor $\tilde{g}(k,s)$ introduced in the generalised SuperEasy Poisson equation~\eqref{eq:gse_poisson} 
(see also equation~\eqref{eq:tildeF}), i.e.,
\begin{equation}
\label{eq:matterpowereasy}
P_{\rm m} (k,s) = f_{\rm cb}^2 \, \tilde{g}^2(k,s) \, P_{\rm cb}(k,s),
\end{equation}
where $\tilde{g}(k,s)$ and $P_{\rm cb}(k,s)$ are, as usual, evaluated at the same wave number~$k$ and time~$s$.

\begin{figure}[t]
    \centering
    \includegraphics[width=0.85\textwidth]{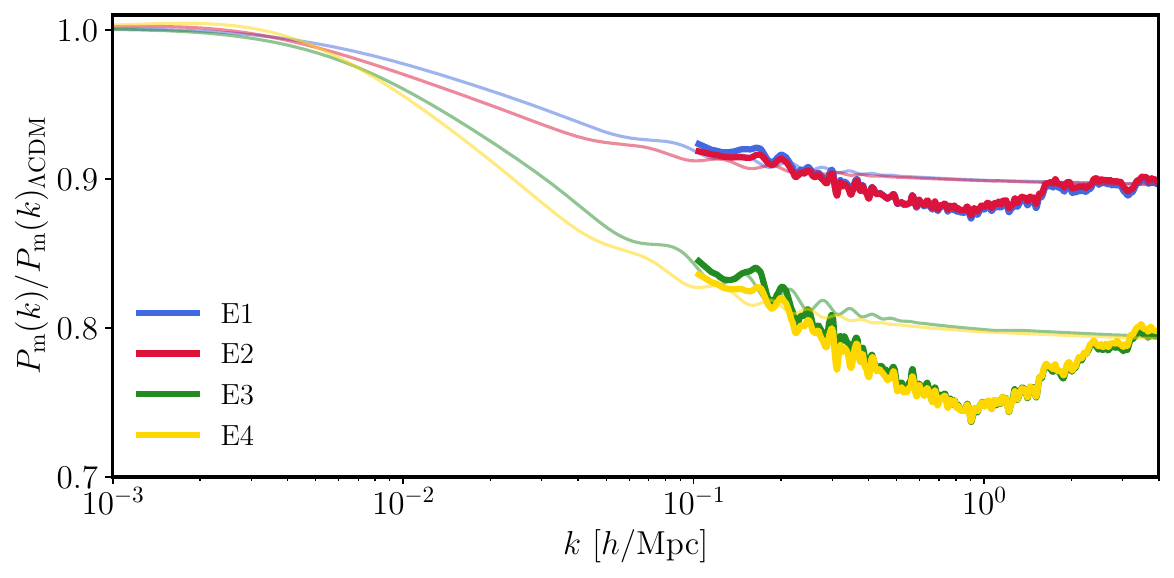}
\caption{Total matter power spectrum at $z=0$ of the mixed HDM models of table~\ref{tab:cosmologies}, normalised to a reference $\Lambda$CDM cosmology with the same cosmological parameters~\eqref{eq:planckparams} and $\Omega_{{\rm hdm},0}=0$. Thick lines depict results from the generalised SuperEasy linear response $N$-body simulations of this work, while thin lines represent the linear theory predictions from~\codo{class}.}
    \label{fig:spoons1}
\end{figure}

Figure~\ref{fig:spoons1} displays the $z=0$ matter power spectra of our four models 
normalised to a $\Lambda$CDM reference specified by the same base cosmological parameters~\eqref{eq:planckparams} but with $\Omega_{{\rm hdm},0}=0$. The thick and thin lines represent, respectively, results from the generalised SuperEasy $N$-body simulations of this work and linear predictions from~\codo{class}. We remind the reader that the models E2 and E4 have been designed such that their $z=0$ {\it linear} total matter power spectra match those of E1 and E3, respectively, on both the large and small scales (the intermediate linear scales cannot be matched because of the different effective free-streaming scales induced by the different effective HDM velocity distributions). Evidently, once this linear matching is in place, non-linear dynamics at $k \gtrsim 0.2 \, h$/Mpc also appears to evolve the total matter power spectra of two different cosmologies to largely the same form at $z=0$. This apparent lack of intrinsic non-linear signature 
immediately suggests that emulators and fitting functions originally designed/calibrated for generating non-linear matter power spectrum corrections for SM neutrino cosmologies can be straightforwardly repurposed for non-standard HDM cosmologies simply by demanding linear matching on non-linear scales.

\begin{figure}[t]
    \centering
    \includegraphics[width=0.9\textwidth]{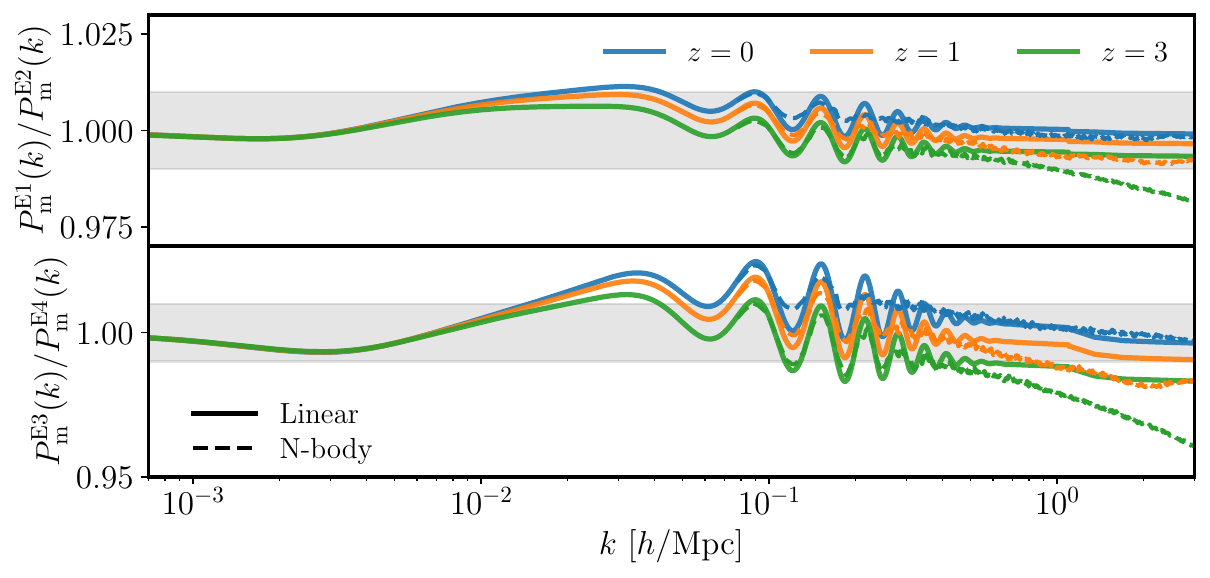}
    \caption{Time evolution of the total matter power spectrum ratios between the models E1 and E2 (top panel) and between the models E3 and E4 (bottom panel). Solid lines show the linear theory output of \codo{class}, while dashed lines represent our results from the linear response $N$-body simulations.}
    \label{fig:matter_z}
\end{figure}

We stop short, however, of claiming that non-standard HDM cosmologies are indistinguishable from SM neutrino cosmologies on non-linear scales. 
Figure~\ref{fig:matter_z} shows the total matter spectrum ratios at $z=0,1,3$ between models E1 and E2, and between models E3 and E4, computed using both linear perturbation theory (dashed) and generalised SuperEasy linear response simulations (solid).  To begin with, we note that there is no simultaneous small-scale linear matching across all redshifts: linear evolution of the total matter density alone, which scales like $\delta_{\rm m}(k) \propto a^{1-(3/5) f_{\rm hdm}}$ in the HDM free-streaming limit during matter domination,%
\footnote{This solution comes from solving ${\rm d}^2\delta_{\rm m} /{\rm d}s^2 - (3/2) (1-f_{\rm hdm} )a^2 {\cal H}^2 \delta_{\rm m}=0$ assuming matter domination.}
must lead to different power spectra for different HDM fractions~$f_{\rm hdm}$.
In fact, once matched at $z=0$, the small-scale mismatches engendered between models E1 and E2 and between models E3 and E4 at higher redshifts can be as large as the mismatches on the intermediate scales. 
Furthermore, non-linear evolution enhances the mismatch at $k \gtrsim 0.2\, h$/Mpc.  Thus, non-standard HDM cosmologies appear to be in principle distinguishable from SM neutrino cosmologies if we are able to measure $P_{\rm m}(k)$ at multiple redshifts.  How well they can be distinguished, however, will depend
ultimately on the sensitivity of an observation to small differences.


\section{Comparison with other HDM simulation methods}
\label{sec:comparison}

References~\cite{Chen:2020kxi} and~\cite{Chen:2022dsv} have previously demonstrated the efficacy of (integrated) SuperEasy linear response, relative to other linear response methods as well as particle HDM simulations, in 
standard massive neutrino cosmologies.  To verify that the generalised SuperEasy approach is similarly valid for mixed HDM cosmologies---especially in cases where the HDM may be significantly colder than SM neutrinos---we perform in this section additional simulations of the same four cosmologies of table~\ref{tab:cosmologies}, using other linear response methods in the literature (section~\ref{sec:compareLRsims}), as well as a particle-based method that goes beyond linearisation (section~\ref{sec:hybrid}).

\subsection{Other linear response simulations}
\label{sec:compareLRsims}

 We consider two linear response approaches, the original SuperEasy linear response~\cite{Chen:2020kxi} and the multi-fluid linear response (MFLR)~\cite{Chen:2020bdf}. These differ from the generalised SuperEasy method of this work only by the modification factor $\tilde{g}(k,s)$ applied to the Poisson equation,     $k^2\Phi(\vec{k},s) =-(3/2) {\cal H}^2 (s) \Omega_{\rm cb}(s) \tilde{g}(k,s)\delta_{\rm cb}(\vec{k},s)$,  to accommodate the presence of HDM:
\begin{equation}
 \tilde{g}(k,s)=\begin{dcases}
             \left(1 + \sum_{i=1}^{N_i}f_{{\rm h}_i} \left[{\cal G}_i(k,s)-1 \right]\right)f^{-1}_{\rm cb}, 
            & ~~~\text{Generalised SuperEasy, any HDM},\\  
            \frac{\left[k+k_{\rm FS}(s)\right]^2}{\left[k+k_{\rm FS}(s)\right]^2-k_{\rm FS}^2(s) f_{\rm hdm}}, & ~~~\text{Original SuperEasy~\cite{Chen:2020kxi}, Fermi-Dirac only},\\
   1+\sum_{i=1}^{N_i}\frac{\Omega_i(s)\delta_{i,\ell=0}(k,s)}{\Omega_{\rm cb}(s)\delta_{\rm cb}(k,s)}, & ~~~\text{Multi-fluid~\cite{Chen:2020bdf}, any HDM},\\
		 \end{dcases} 
   \label{eq:tildeF}
\end{equation} 
where the various quantities are defined below. Our modified version of \codo{gadget-4} incorporates all of the above three options.\footnote{A runtime option can be specified in the \codo{param.txt} file to choose between the three linear repsonse methods: \codo{NLR}=1,2,3 for original SuperEasy, MFLR, and generalised SuperEasy, respectively.} 
We note that there exist also other linear response methods in the literature not considered in this work, such as the method presented in reference~\cite{Ali-Haimoud:2012fzp}, which is equivalent to the MFLR approach of reference~\cite{Chen:2020bdf} when the latter is run with sufficient momentum resolution.


\paragraph{Original SuperEasy.}
We have already discussed the original SuperEasy method in section~\ref{sec:ose}, where the integrated free-streaming scale $k_{\rm FS}(s)$ is given in equation~\eqref{eq:intfs}, and $f_{\rm hdm}$ is simply the total HDM fraction.
Strictly speaking, the expression~\eqref{eq:intfs} applies only to an effective HDM whose homogeneous background phase space density is described by a relativistic Fermi-Dirac distribution.  This condition is certainly not borne out by models E1 and E3 (which contain QCD axions and thermal bosons, respectively), but can be reasonably satisfied by the neutrino-only models E2 and E4 according to figure~\ref{fig:psd}, despite their non-degenerate particle mass spectra.  

We therefore apply the original SuperEasy linear response only to models E2 and E4, using the integrated free-streaming scale of equation~\eqref{eq:intfs} with $m_\nu=m_{\rm hdm}$ and $T_{\nu,0}=T_{\rm hdm}$ given in table~\ref{tab:cosmologies}.  The simulations are then conducted in the same way as in reference~\cite{Chen:2020kxi} but with an initialisation redshift of $z=99$.


\paragraph{Multi-fluid linear response (MFLR).}

The MFLR approach~\cite{Chen:2020bdf}, itself based upon the perturbation theory developed in~\cite{Dupuy:2013jaa,Dupuy:2014vea}, partitions an HDM fluid into $N_i$ flows, each labelled by the flow's initial comoving momentum magnitude $\tau_i$ and weighted by its reduced energy density $\Omega_i(s)$.
Each flow is further decomposed into Legendre multipole moments at multipoles $\ell = 0, \cdots N_\mu-1$, according to the alignment of the flow momentum to the wave vector, $\mu=\hat{k} \cdot \hat{\tau}_i$.  The fluid density and velocity perturbations $\{\delta_{i, \ell}, \theta_{i, \ell}\}$ are then tracked by $N_i \times N_\mu$ sets of continuity and Euler equations.
Only the density monopole of each flow, $\delta_{i,\ell}$, contributes to the total HDM energy density perturbation, i.e., $\delta_{\rm hdm}= (1/\Omega_{\rm hdm}) \sum_{i=1}^{N_i} \Omega_i \, \delta_{i,\ell=0}$. Since it involves no interpolation, MFLR might be considered ``exact'' in comparison with SuperEasy methods.  In practice, however, to minimise computational cost some approximation must be invoked in the method's actual implementation in an $N$-body code (see below).

As in reference~\cite{Chen:2020bdf} we have interfaced our MFLR module with \codo{gadget-4}.%
\footnote{A stand-alone version of our MFLR code is publicly available at {\tt \href{https://github.com/upadhye/MuFLR-HDM}{github.com/upadhye/MuFLR-HDM}}~.}  
After an initialisation phase between $z=999$ and $z=99$, in which the MFLR module is run for purely linear perturbations in order to find the attractor solutions for the fluid perturbations, $N$-body evolution is switched on for the cold matter at $z=99$.  From this point onwards, at each time step the MFLR module takes as an input from \codo{gadget-4} a {\it directionally-averaged} $k$-space gravitational potential $\Phi(k,s) \equiv \langle \Phi(\vec{k},s)\rangle_{\hat{k}}$, and returns as output the HDM-to-cold matter density perturbation ratio, $\Omega_{\rm hdm}\delta_{\rm hdm}/\Omega_{\rm cb} \delta_{\rm cb}$.   In other words, the MFLR module tracks only the directionally-averaged HDM perturbations and discards some phase information in the process.
The ratio goes in turn into the factor $\tilde{g}(k,s)$ of equation~\eqref{eq:tildeF}, which modifies the $k$-space Poisson equation on the \codo{gadget-4} PM grid, assuming the directionally-averaged HDM perturbation to be locally in phase with the cold matter perturbations at $\vec{k}$.

The implementation~\cite{Chen:2020bdf} partitioned the HDM fluids into $50$ equal-number density bins.  Here, as in the companion paper~\cite{Flows2}, we improve upon the binning scheme by adopting the method of Gauss-Laguerre quadrature.  Because of the correspondence between Eulerian and the semi-Lagrangian MFLR (see appendix~\ref{app:SuperEasytoMFLR}), the same Gauss-Laguerre momentum binning strategy discussed in section~\ref{sec:LRsims} and displayed in figure~\ref{fig:psd2} applies to the MFLR flow discretisation, and we again adopt $N_i=15$.  Reference~\cite{Chen:2020bdf} showed that the truncation error in the Legendre expansion scales as $N_\mu^{-2}$, which motivates us to choose $N_\mu = 10$ for percent-level accuracy in the angular expansion.

\bigskip

\begin{figure}[t]
\begin{center}
    \includegraphics[width=0.49\textwidth]{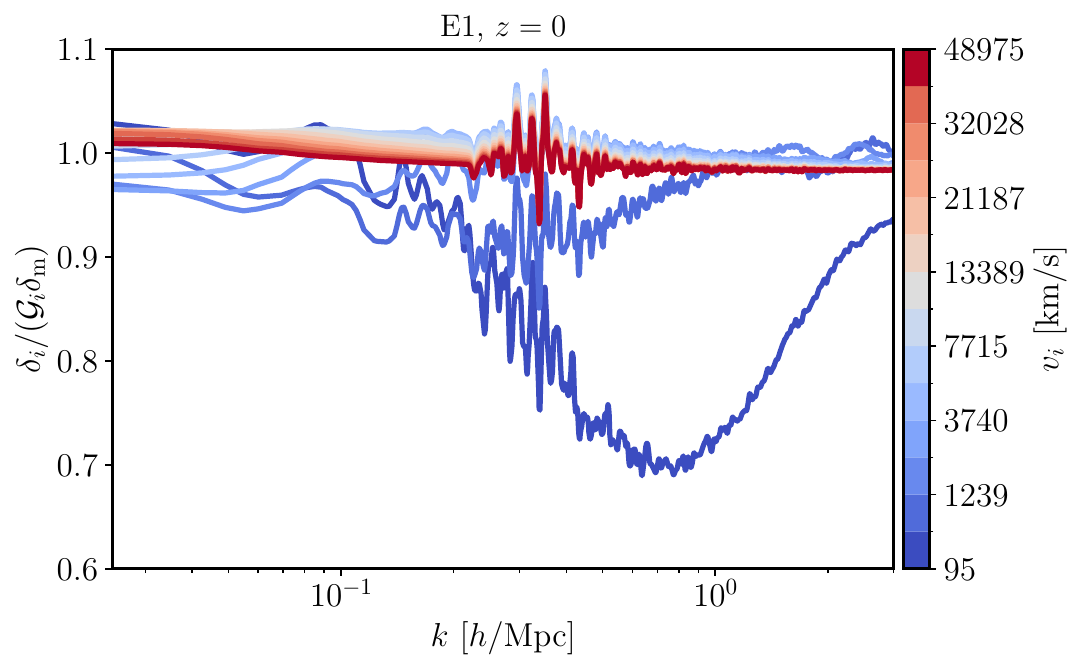}
    \includegraphics[width=0.49\textwidth]
    {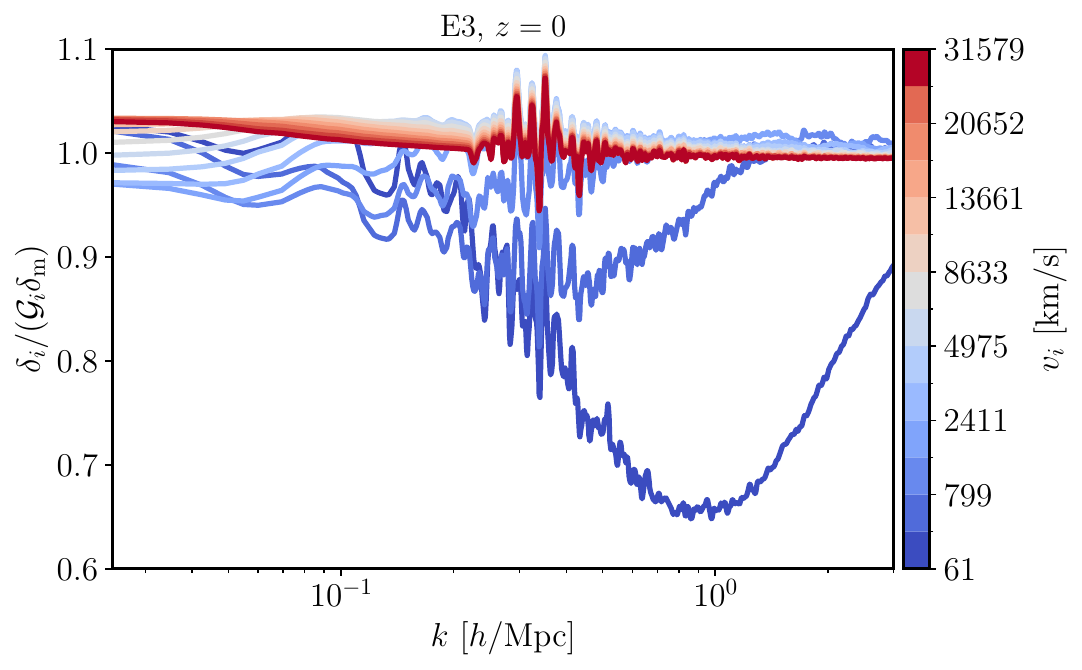}
    \caption{Comparison from N-body simulations of the generalised SuperEasy approach and the MFLR approach in their predictions of the ratio $\delta_i/\delta_{\rm m}$, where $\delta_i$ is the HDM density contrast in the momentum bin~$i$, and $\delta_{\rm m}$ is the total matter density contrast. 
    In the generalised SuperEasy method, this ratio is simply given by the interpolation function~\eqref{eq:interpoliationG}, i.e., ${\cal G}_i(k)={\cal G}(k,p_i)$, evaluated the bin momentum~$p_i$.  {\it Left}: The MFLR-to-SuperEasy prediction of $\delta_i/\delta_{\rm m}$ at $z=0$ for model E1.  {\it Right}: Same as left but for model E3.
         }
    \label{fig:mf_gse_check}
    \end{center}
\end{figure}

Figure~\ref{fig:mf_gse_check} compares the ratio of the each MFLR flow's monopole density perturbations to the total matter perturbation, i.e., $\delta_{i, \ell=0}/\delta_{\rm m}$, with the corresponding prediction of the generalised SuperEasy approach, namely, $\delta_i/\delta_{\rm m}= {\cal G}_i$, where ${\cal G}_i(k,s)={\cal G}(k,p_i,s)$ is the interpolation function~\eqref{eq:interpoliationG} evaluated at the momentum of bin~$i$.  We show results for models E1 and E3 at $z=0$, both of which have non-standard HDM contents.  Clearly, the agreement between the generalised SuperEasy approach and the MFLR method is very good especially for the faster flows, where sub-10\% concordance is seen across the entire simulated $k$-range, consistent with the estimates of section~\ref{sec:gse}.  In both models the slowest flow shows the largest discrepancy: a maximum of around 30\% centred, interestingly, on the flow's free-streaming scale $k_{{\rm FS},i}$.  Notwithstanding the approximations employed in the MFLR computation (such that its results are strictly not ``exact'' within the linear response class of methods), this discrepancy trend strongly suggests that the interpolation function~\eqref{eq:interpoliationG} itself may be in need of further refinement, a subject we leave for future work.

\begin{figure}[t]
    \includegraphics[width=\textwidth]{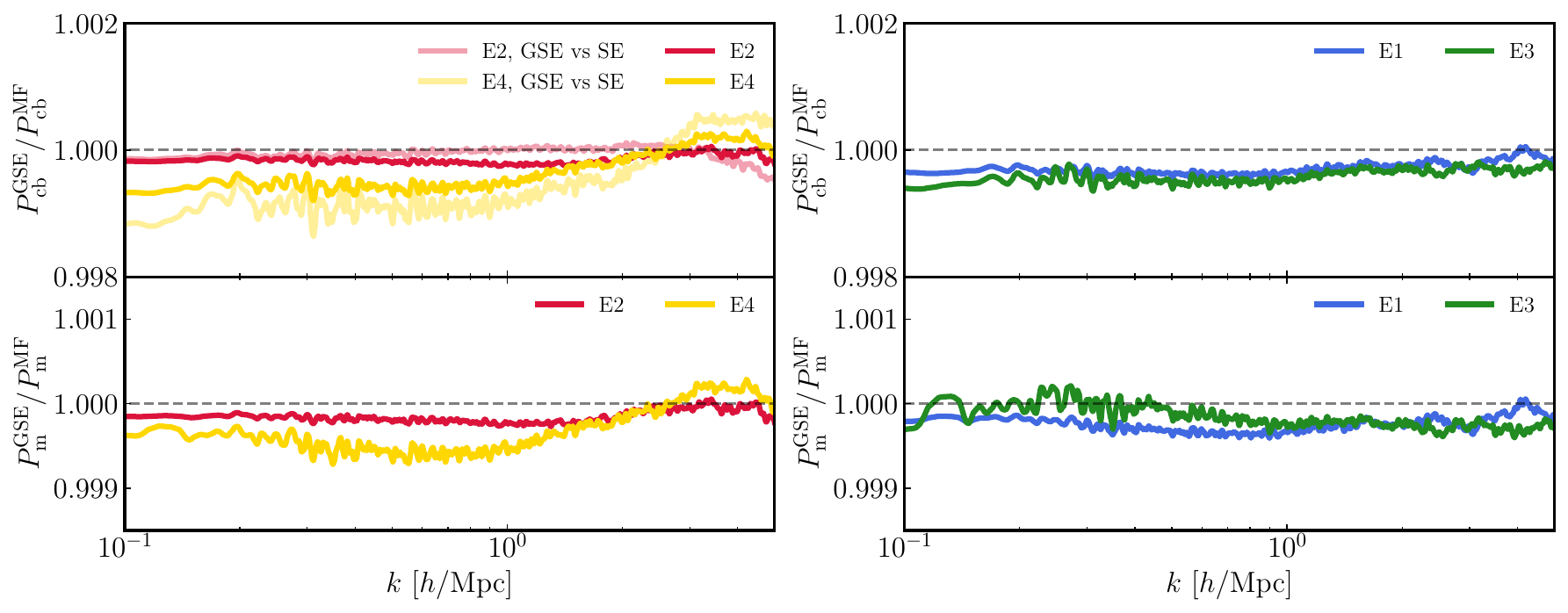}
    \caption{\emph{Top}: Comparison of the generalised SuperEasy linear response simulation of the $z=0$ cold matter power spectrum $P_{\rm cb}^{\rm GSE}(k)$ to its MFLR counterpart $P^{\rm MF}_{\rm cb}(k)$ and, where applicable, the original SuperEasy linear response (shaded lines). On the left panel we show power spectrum ratios for models E2 and E4, which contain only three SM neutrinos and whose effective distribution can be approximated as a relativistic Fermi-Dirac distribution (see figure~\ref{fig:psd2}) such that the original SuperEasy method~\cite{Chen:2020kxi} applies, while on the right panel we display ratios for models E1 and E3 containing an admixture of axions/bosons and SM neutrinos; the original SuperEasy method does not apply to these models.  In all four cosmologies, we find $\sim 0.1$\% agreement between all three linear response methods across the whole simulation $k$-range. \emph{Bottom}: Same as  the top row, but for the total matter power spectrum $P_{\rm m}(k)$, constructed from $P_{\rm cb}(k)$ via the relation $\delta_{\rm m}(k)=f_{\rm cb}\, \tilde{g}(k)\, \delta_{\rm cb}(k)$.
}
    \label{fig:gse_mf_nbody}
\end{figure}

In terms of predictions of the cold matter power spectrum,  the top row of figure~\ref{fig:gse_mf_nbody} shows $P_{\rm cb}(k)$ at $z=0$ of models E1--E4 computed with the generalised SuperEasy method, in comparison with the predictions of MFLR and, in the case of models E2 and E4, the predictions of the original SuperEasy linear response as well (left panel).
On the bottom row we show the same ratios but for the total matter power spectrum~$P_{\rm m}(k)$.
Evidently, for both observables, variations between methods are typically within $\sim 0.1\%$ across the whole simulation $k$-range, with the smaller HDM masses (or, equivalently, smaller HDM energy fractions $f_{\rm hdm}$) exhibiting better agreement at $k \lesssim 1\, h$/Mpc. Considering the excellent concordance---and in spite of the different approximations inherent in them---we can conclude that, within the linear response class of HDM simulations, the generalised SuperEasy method of this work is an extremely simple and efficient way to model a non-standard or effective HDM population that also comes with essentially no runtime penalties compared to a standard $\Lambda$CDM simulation.


\subsection{Hybrid multi-fluid\texorpdfstring{\Plus{}}{+}particle simulations}\label{sec:hybrid}

Having tested generalised SuperEasy linear response against other linear response methods, we now turn to the next question: is linear response, which explicitly linearises the HDM equation of motion, adequate for the modelling of HDM in the non-linear regime?  This question has been addressed in the context of degenerate SM neutrinos in references~\cite{Bird:2018all,Chen:2022dsv}.  As a rule of thumb, linear response methods can predict the total HDM density contrast at $k \lesssim 0.5\, h$/Mpc to $\lesssim 5$\% accuracy, if the particles' Lagrangian velocities $v$ do not exceed $\sim 600$~km/s. If, however, we are only interested in the clustering statistics of the cold matter and the total matter, then linear response can be accurate to $\lesssim 0.5$\% across the whole $k$-range of interest for neutrino mass sums not exceeding $\sum m_\nu \simeq 0.5$~eV.

We expect these conditions to largely hold for the mixed HDM scenarios considered in this work.  Nonetheless, because the E1 and E3 models contain bosonic HDM components that are (i)~colder and (ii)~skewed towards the low-momentum tail (see figures~\ref{fig:psd} and~\ref{fig:psd2}), one would generically expect these cosmologies to develop more non-linearities in the HDM sector than SM neutrinos. To study how these non-linearities develop and what kind of impact they have on the cold matter and total matter spectra, we adapt the Hybrid MFLR\Plus{}particle $N$-body approach of reference~\cite{Chen:2022dsv} to simulate the mixed HDM models of this work.

The Hybrid approach extends MFLR by systematically converting HDM flows in the low-velocity tail to a particle representation, while retaining the high-velocity flows on the linear response description~\cite{Chen:2022dsv}.   A converted flow $i$ with an energy density of $\Omega_i$ is represented by $N^{\rm hdm}_{\rm part}$  $N$-body particles placed on real-space grid sites $\vec{x}$.  Each particle has a mass $M(\vec{x})$ determined by
\begin{equation}
    M(\vec{x})=\langle M\rangle\left[1+\delta_{i} (\vec{x}, z_c)\right], ~~~~~\langle M\rangle=\frac{3H^2_0L^3\Omega_{i}}{8\pi G N^{\rm hdm}_{\rm part}}
\end{equation} 
at the conversion redshift $z_c$, where $\langle M \rangle$ is the average $N$-body particle mass, 
and $\delta_{i} (\vec{x}, z_c)$ is the real-space density fluctuation at the grid site.%
\footnote{Using different particle masses to represent the initial HDM fluctuations bypasses the need to displace particles from grid sites at initialisation, in contrast to what is commonly done to (equal-mass) cold particles.} 
In turn, the real-space density fluctuation $\delta_{i}(\vec{x},z_c)$ at the grid sites are constructed from
\begin{equation}
    \delta_{i} (\vec{x}, z_c) =\mathscr{F}^{-1}[\delta_{i, \ell=0} (k, z_c)e^{i\varphi(\vec{k}, z_c)}]
\end{equation}
in the monopole ($\ell=0$) approximation,%
\footnote{Removing the $\ell>0$ multipoles simplifies the picture purely for computational reasons. As discussed in~\cite{Chen:2022dsv}, this simplification can lead to substantial unphysical transients around the HDM free-streaming scale if the simulation is initialised very late. To limit this issue we initialise the HDM particles as early as possible.} 
where $\mathscr{F}$ denotes the Fourier transform operator, $\delta_{i,\ell=0}(k,z_c)$ is the $\ell=0$ MFLR solutions at the conversion redshift $z_c$, and the phase,
\begin{equation}
    e^{i\varphi(\vec{k}, z_c)}=\frac{\Phi(\vec{k}, z_c)}{\sqrt{\langle\vert\Phi(\vec{k}, z_c)\vert^2\rangle}}, 
\end{equation}
is formed from the gravitational potential $\Phi(\vec{k},z_c)$ due to all particles---cold ones and any previously converted hot ones---already in the simulation at the time of conversion.  In addition, each HDM  $N$-body particle receives a velocity kick $\vec{u}_{i}(\vec{x}, \vec{\tau}_{i},z_c)$ at conversion given by 
\begin{equation}
    \vec{u}_{i}(\vec{x}, \vec{\tau}_{i},z_c) =\frac{1}{a(z_c) m_{\rm hdm}}\left(\vec{\tau}_{i}-i{\cal F}^{-1}\left[\frac{\vec{k}}{k^2}\theta_{i, \ell=0}(k,z_c)e^{i\varphi(\vec{k}, z_c)}\right]\right),
\end{equation}
with $\vec{\tau}_i=|\tau_i| \hat{\tau}_i$ determined from the flow speed $\vert\tau_{i}\vert/m_{\rm hdm}$ and a random direction $\hat{\tau}_{i}$ sampled from $4\pi$ solid angle.
We implement this HDM particle initialisation procedure in our version of~\codo{gadget-4} using \codo{ngenic}, the built-in initial condition generator module.  Note that because the procedure computes the phases of the existing potentials on the Fourier grid, it is necessary that CDM particles be present in the snapshot prior to placing HDM particles.

\begin{figure}[t]
    \centering
    \includegraphics[width=0.93\textwidth]{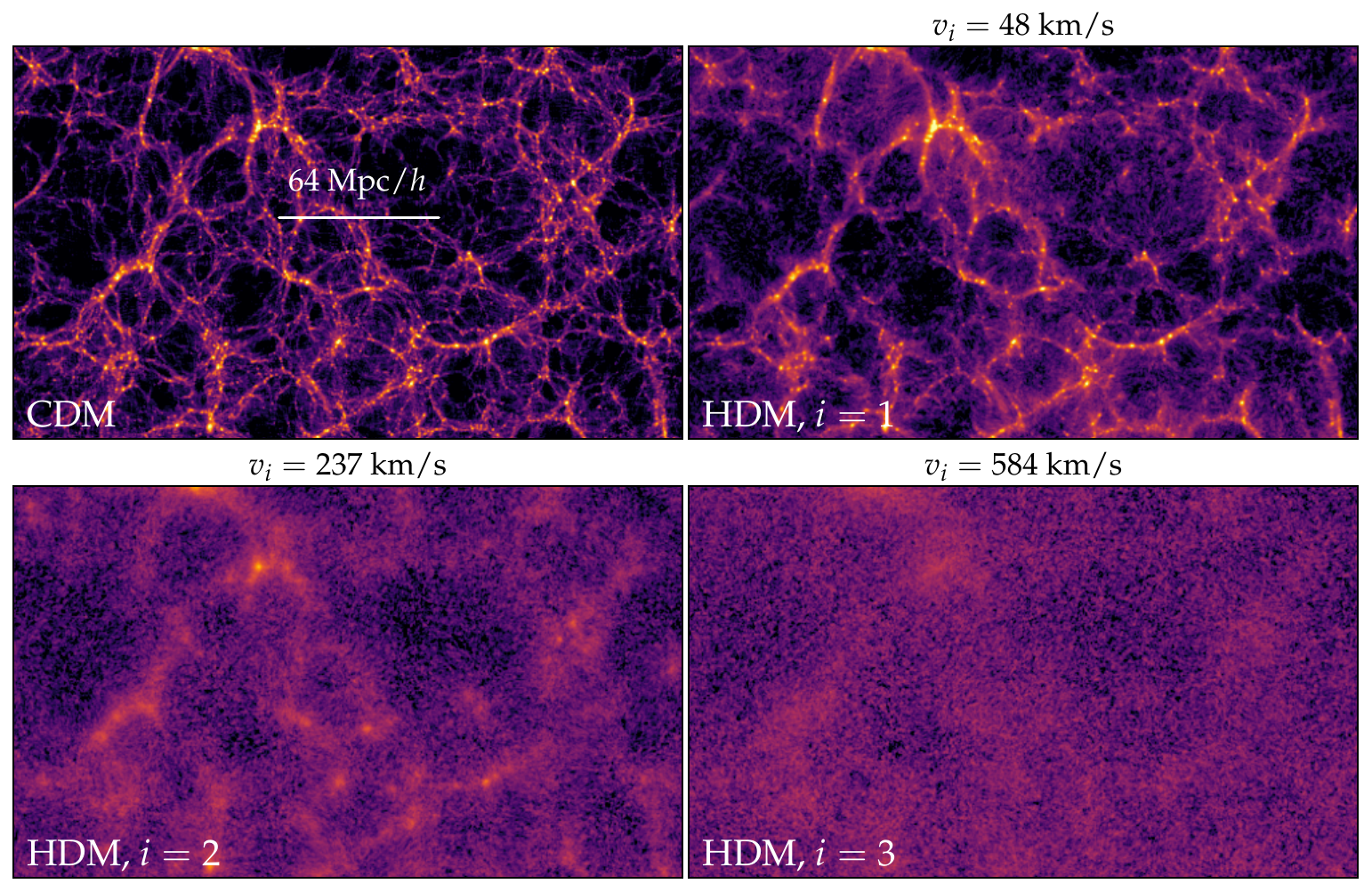}
    \caption{3D-to-2D projection of the density contrast field $\delta$ at redshift $z=0$ from a Hybrid simulation of model E4, obtained from a 20~Mpc$/h$-thick slice of the simulation box integrated along the thickness.    
    We show the density contrast in the cold particles (CDM) and in the HDM particles from the converted flows $i=1,2,3$. }
    \label{fig:proj}
\end{figure}

Following the $v\lesssim 600$~km/s guideline~\cite{Bird:2018all,Chen:2022dsv}, we convert the $i=1, 2$ flows for models E1--E3 and the $i=1, 2, 3$ flows for model E4 (see also figure~\ref{fig:psd2}).  Each converted flow is represented by $N^{\rm hdm}_{\rm part}=512^3$ $N$-body particles. To avoid transients~\cite{Chen:2022dsv}, the first conversion is performed directly after the initialisation of the cold particles, i.e, at $z=99$, while the last conversion occurs at $z_c=89$.
Figure~\ref{fig:proj} shows the 3D-to-2D projections of the cold matter and converted HDM (flows $i=1,2,3$) density contrast fields of model E4 at $z=0$ from a Hybrid simulation.  The difference in clustering between the different flows is immediately evident, with the structure of the $i=1$ strongly resembling the clustering of the CDM, $i=3$ being much more diffuse, and $i=2$ somewhere in-between.

\begin{figure}[t]
    \centering
       \includegraphics[width=0.48\textwidth]{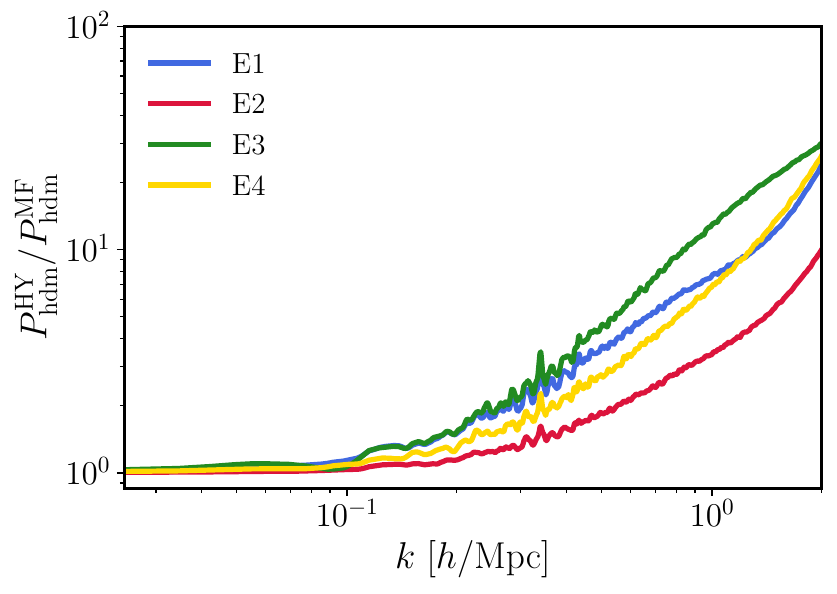}
    \includegraphics[width=0.48\textwidth]{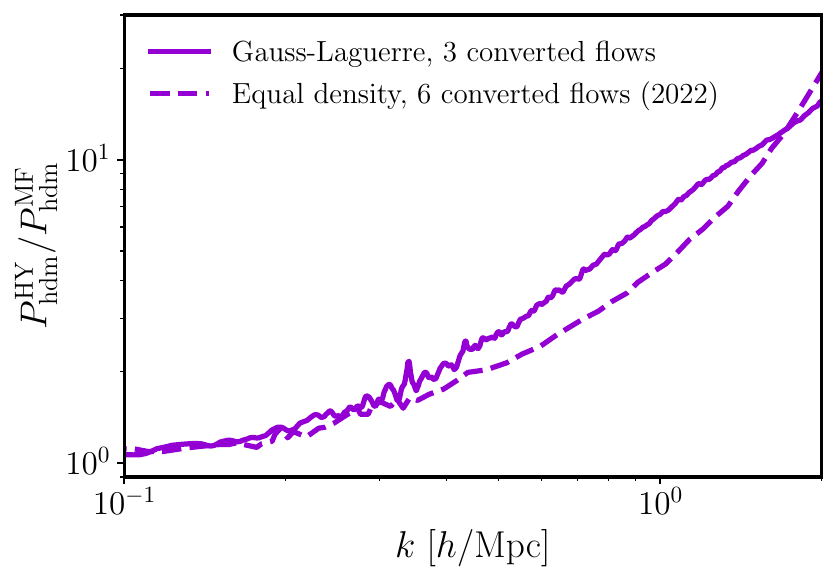} 
    \caption{Total HDM power spectrum estimated from the Hybrid approach $P^{\rm HY}_{\rm hdm}(k)$ relative to its MFLR counterpart $P^{\rm MF}_{\rm hdm}(k)$.  The ratio shows the non-linear enhancement arising from conversion of a subset of flows to a particle representation. 
 {\it Left}: Non-linear enhancement seen in the four cosmological models E1--E4 of table~\ref{tab:cosmologies}. {\it Right}: Non-linear enhancement in a model consisting of three degenerate SM neutrinos with $\sum m_{\nu}=0.465$~eV considered in reference~\cite{Chen:2022dsv}.  The equal-number density binning scheme used~\cite{Chen:2022dsv} (2022, dashed line) visibly underestimates the enhancement on scales $k\gtrsim 0.5 \, h/$Mpc relative to the Gauss-Laguerre scheme (solid line) advocated in this work.}
    \label{fig:ahdm_hybrid}
\end{figure}

The left panel of figure~\ref{fig:ahdm_hybrid} shows the total HDM power spectra of models E1--E4 from the Hybrid simulations, defined as
\begin{equation}
    P^{\rm HY}_{\rm hdm}(k)=\left\vert f^{\rm HY}_{\rm hdm}\sqrt{P^{\rm hdm}_{\rm gadget}(k)}+\sum_{i=i^*}^{N_i}f_{{\rm h}_i}\delta_{i, \ell=0}\right\vert^2,
\end{equation} 
where $P^{\rm hdm}_{\rm gadget}(k)$ is the power spectrum of \emph{all} HDM particles (i.e., the converted flows) in the simulation box, which have total a density fraction $f_{\rm hdm}^{\rm HY}=\sum_{i=1}^{i^*-1}f_{{\rm h}_i}$, and $i^*$ is the first unconverted flow. This expression carries the assumption that the unconverted flows are in phase with the sum of the converted flows, which should be a reasonable approximation.
As in~\cite{Chen:2022dsv}, we compare these Hybrid HDM power spectra with their counterparts obtained from MFLR simulations, given by \begin{equation}
    P^{\rm MF}_{\rm hdm}(k)=\left\vert\sum_{i=1}^{N_i} f_{{\rm h}_i}\delta_{i, \ell=0}\right\vert^2.
    \label{eq:matter_mf}
\end{equation} 
Evidently, all four models see significant non-linear enhancement in the total HDM power at $k \gtrsim 10^{-1}\, h$/Mpc.  At $k\simeq 1\, h$/Mpc, the minimum enhancement across the four mixed HDM cosmologies is a factor of~$\sim 3$.  In particular, model E3, which contains a thermal boson, 
shows the largest increase in power--- a factor of $\sim 13$---because of the dominant low-$p$ tail of the effective HDM distribution (see figure~\ref{fig:psd2}).

Before moving on, let us also comment on the Gauss-Laguerre binning scheme of this work versus the equal-number density bins used in reference~\cite{Chen:2022dsv}.  The right panel of figure~\ref{fig:ahdm_hybrid} shows the total HDM power spectrum ratio $P_{\rm hdm}^{\rm HY}/P_{\rm hdm}^{\rm MF}$ for an HDM model consisting of three degenerate SM neutrinos of total mass 
$\sum m_{\nu}=0.465$ eV computed in reference~\cite{Chen:2022dsv} using the equal-number density binning scheme, versus a new hybrid simulation using Gauss-Laguerre quadrature.  Firstly, because the slowest particle speed in the Gauss-Laguerre scheme is a factor of $\sim 7.5$ smaller than in the equal-number scheme, the former is more able at capturing the non-linearites in the low-$p$
tail of the neutrino distribution, resulting in $\sim 40\%$ more total HDM power at $k\simeq 1\, h$/Mpc (and bringing the prediction more in line with the estimates of, e.g.,~\cite{Elbers:2020lbn}).
Secondly, the Gauss-Laguerre scheme requires only three (out of 15) MFLR flows to be converted to a particle representation, in contrast to six (out of 20) flows in the equal-number scheme---which turns out to underestimate the power anyway. The companion paper~\cite{Flows2} discusses this issue in more detail, but suffice it to say here that the Gauss-Laguerre binning scheme is clearly the more efficient of the two.

\begin{figure}
    \centering
    \includegraphics[width=\textwidth]{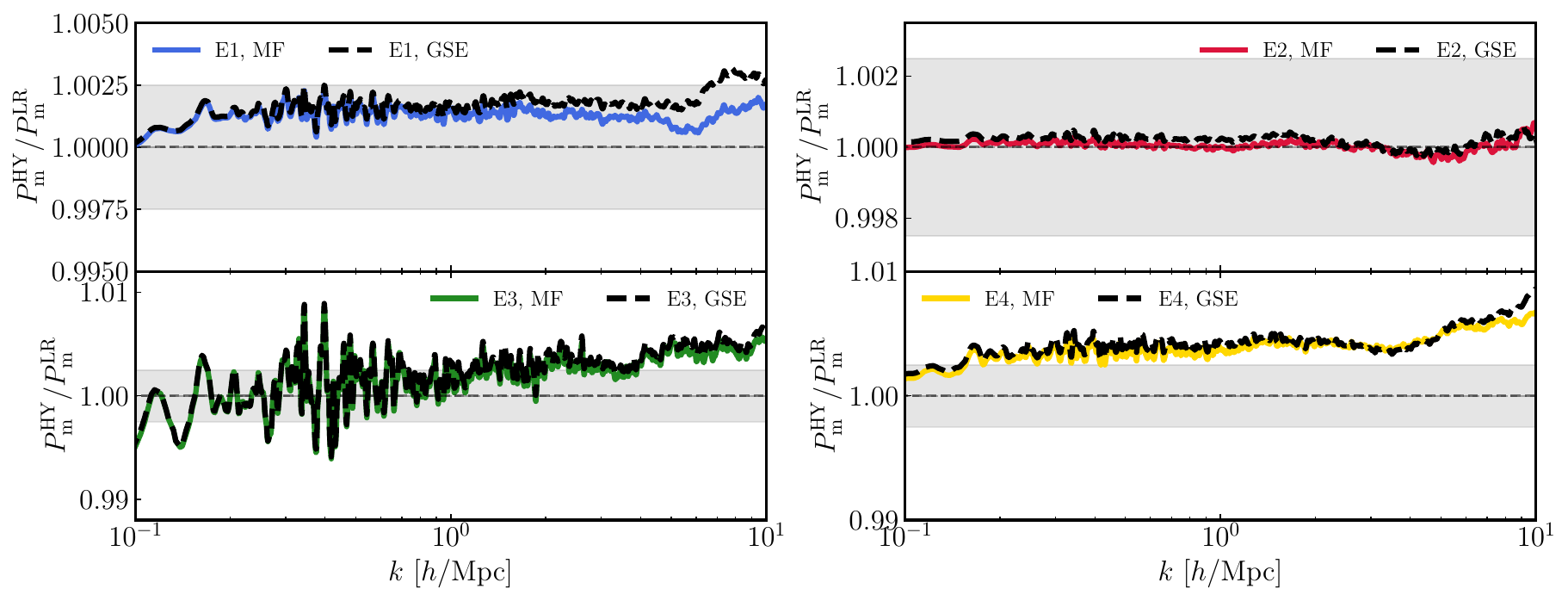}
    \caption{Hybrid to linear response total matter power spectrum ratio for our four HDM models at $z=0$. We consider MFLR (coloured solid) and generalised SuperEasy linear response (GSE; black dashed lines), both of which show a sub-percent level accuracy relative to the Hybrid runs. The Hybrid and the linear response power spectra have been estimated from the relations~\eqref{eq:matter_hyb_gadget} and~\eqref{eq:matterpowereasy}, respectively, using in the latter case the $\tilde{g}(k,s)$ functions~\eqref{eq:tildeF} and $P_{\rm cb}(k)$ extracted from \codo{gadget-4}.
    We highlight in grey the $\pm 0.25$\% region around unity. 
    }
    \label{fig:matter_hyb}
\end{figure}

Lastly, we consider the impact of HDM non-linearities on the total matter spectrum, computed from a Hybrid simulation as
\begin{equation}
    P^{\rm HY}_{\rm m}(k)=\left\vert \left(f_{\rm cb}+\sum_{i=1}^{i^*-1} f_{{\rm h}_i}\right) \sqrt{P_{\rm gadget}(k)} 
    +\sum_{i=i^*}^{N_i}f_{{\rm h}_i}\delta_{i, \ell=0}
    \right\vert^2,\label{eq:matter_hyb_gadget}
\end{equation} 
where $P_{\rm gadget}(k)$ is the power spectrum of {\it all} particles, both cold and hot.  Figure~\ref{fig:matter_hyb}~shows the total matter power spectra constructed in this manner for the four models E1--E4 under consideration, normalised to their corresponding MFLR estimates.  Evidently, because of the subdominant nature of HDM, the difference between the two estimates of $P_{\rm m}(k)$ is small across the four models: in the case of models E3 and E4, the maximum discrepancy at $k \lesssim {\cal O}(1)\, h$/Mpc is of order 0.25\%; for E1 and E2, the discrepancy is sub-0.25\%.  Thus, while the HDM non-linearities do have a strong impact on the HDM clustering {\it itself}, ultimately these non-linearities translate only into a small effect in the total matter density perturbations.  We conclude on this basis that HDM linear response is a sufficient method for estimating the {\it total matter} density perturbations in subdominant HDM cosmologies to sub-1\% accuracy---even for those colder HDM variants that are already disfavoured by current cosmological data (e.g., E3 and E4).

\section{Conclusions}
\label{sec:conclusions}

We have investigated in this work analytical and numerical techniques to account for the presence of hot dark matter in cosmological $N$-body simulations. Specifically, we have generalised the titular SuperEasy linear response method, developed originally for standard-model neutrinos~\cite{Chen:2020kxi}, to arbitrary momentum distributions, including models with non-degenerate neutrino masses as well as mixed HDM models containing additional non-standard thermal relics such as thermally-produced QCD axions. 
The method employs approximate analytic solutions of the HDM phase space perturbation in response to a CDM density field in the clustering limit ($k \ll k_{{\rm FS},p}$) and the free-streaming limit ($k \gg k_{{\rm FS},p}$), and interpolates between them using a simple rational function in $k/k_{{\rm FS},p}$ that does not require any fitting. Integration over momentum $p$, weighted by the effective HDM momentum distribution, then completes the estimate of the HDM density fluctuation.  Because the free-streaming wave number $k_{{\rm FS},p}$ is algebraic in the effective HDM mass, the total matter density, and the scale factor of interest, the method can be immediately and very simply implemented into any Particle-Mesh simulation code as a modification to the Fourier-space gravitational potential. This in turn tricks the cold particles into trajectories as if HDM particles were present in the simulation box.

We have tested this generalised SuperEasy linear response approach against existing linear response methods: the original SuperEasy method~\cite{Chen:2020kxi} which applies only to relics following a relativistic Fermi-Dirac distribution, and the multi-fluid linear response~\cite{Chen:2020bdf} which partitions the HDM into multiple flows and evolves a set of fluid equations for each flow.  We find better than $0.1$\% agreement across all three methods in their predictions of the total matter and cold matter power spectra for a range of mixed HDM cosmologies specified in table~\ref{tab:cosmologies} across the whole simulated $k$-range.  This indicates that the generalised SuperEasy linear response approach is at least as accurate as any other linear response method, and is extremely fast and simple to implement to boot.

In addition, we have tested the validity of linear response methods {\it themselves}, by performing Hybrid multi-fluid\Plus{}particle simulations~\cite{Chen:2022dsv} of the same cosmologies, which give the slowest fraction of the HDM population a particle realisation---to capture the full extent of non-linearities---while retaining the rest on multi-fluid linear response.  Following~\cite{Chen:2022dsv} we convert to a particle representation the section of HDM population with present-day velocities satisfying $v<600$ km$/$s. But, improving upon~\cite{Chen:2022dsv}, we use the Gauss-Laguerre binning scheme to better sample the low-$p$ tail of the HDM distributions (instead of the equal-number density binning scheme of ~\cite{Chen:2022dsv}, which turns out to underestimate the total HDM power by $\sim 40\%$).  We find that non-linear HDM clustering enhances the total {\it HDM power spectrum} at $k\simeq 1\, h$/Mpc by factors between $\sim 3-13$ relative to the linear response prediction.   However, owing to the small HDM energy density relative to the total matter density, this non-linear enhancement ultimately does not translate to more than a $\sim 0.25\%$ increase in the {\it total matter power spectrum}.  We can therefore conclude linear response methods suffice to model subdominant HDM cosmologies to sub-1\% accuracy, even for those non-standard, colder relics like QCD axions that are already disfavoured by current cosmological data.

Finally, we have examined the non-linear signatures of non-standard HDM cosmologies.  Comparing a mixed neutrino\Plus{}axion model to a pure neutrino model tuned to match the $z=0$ total {\it linear} matter power spectrum on the very large and very small scales, we find that non-linear evolution preserves the matching at $z=0$.  The same trend is also seen for the mixed neutrino\Plus{}boson model.   This indicates that there is no intrinsically different non-linear signature between HDM cosmologies. However, matching cannot be simultaneously achieved across all redshifts, because matching small-scale suppression of two HDM cosmologies generally leads to a mismatch in the HDM fractions.  This in turn leads to different redshift evolution between different HDM cosmologies even in linear perturbation theory, and non-linear evolution serves to augment the difference.
We therefore conclude that observations of the large-scale structure distribution at multiple redshifts will be necessary to distinguish a standard neutrino HDM cosmology from a non-standard one.


\section*{Acknowledgements}

We thank Alessio Notari and Fabrizio Rompineve for discussions and for providing the distributions of thermal axions. This research was undertaken using the HPC systems \emph{Gadi} from the National Computational Infrastructure (NCI) supported by the Australian Government, and \emph{Katana} at UNSW Sydney. MM acknowledges support from the DFG grant LE 3742/8-1.  GP and Y$^3$W are supported by the Australian Government through the Australian Research Council’s Future Fellowship (FT180100031) and
Discovery Project (DP240103130) schemes. 

\appendix

\section{Connection to multi-fluid linear response}
\label{app:SuperEasytoMFLR}

The semi-Lagrangian, multi-fluid formulation~\cite{Dupuy:2013jaa,Dupuy:2014vea} partitions a hot dark matter population into multiple flows, where the $\alpha$-th flow is labelled by its initial comoving momentum $\vec{\tau}_{\alpha}$, also called the Lagrangian momentum.  A set of fluid equations is used to describe the density perturbation $\delta_{\alpha}(\vec{k},\vec{\tau}_\alpha,s)$ and momentum divergence $\theta_{\alpha}^P(\vec{k},\vec{\tau}_\alpha,s)$ of the $\alpha$-th flow.  In the non-relativistic, sub-horizon limit, the fluid equations take the form~\cite{Chen:2020bdf}
\begin{equation}
\begin{aligned}
\frac{{\rm d} \delta_\alpha}{{\rm d} s} =& -{\rm i} \frac{k \mu \tau_\alpha}{m} \delta_\alpha - \frac{1}{m} \theta^P_\alpha, \\
\frac{{\rm d} \theta^P_\alpha}{{\rm d} s} =& -{\rm i} \frac{k \mu \tau_\alpha}{m} \theta^P_\alpha + a^2 m\,  k^2 \Phi,
\label{eq:mflrFluidEqn}
\end{aligned}
\end{equation}
where $\mu \equiv \hat{k} \cdot \hat{\tau}_\alpha$.  Decomposing $X(\mu) = \sum_{\ell=0}^\infty (-{\rm i})^\ell P_\ell(\mu) X_\ell$, where $P_\ell$ is a Legendre polynomial of degree $\ell$, it was previously shown in reference~\cite{Chen:2022dsv} that $\delta_{\alpha,\ell}$ has a formal solution
\begin{equation}
\label{eq:thetaPla}
 \delta_{\alpha,\ell} (\vec{k},s)   =  - (2 \ell +1) \, k^2  \int_{s_{\rm i}}^s {\rm d} s' \, a^2 (s') \, (s-s')\, \Phi(\vec{k},s') \,  j_\ell  \left[k \tau_\alpha (s-s')/m\right],
\end{equation}
with spherical Bessel function $j_\ell(q)$.

We are interested in the monopole $\ell=0$, as only the monopole contributes to the gravitational potential $\Phi$.  Specifically, $k^2 \Phi  = -(3/2) {\cal H}^2(s) [\Omega_{\rm cb}(s) \delta_{\rm cb}+ \sum_{\alpha =1}^{N_\tau} \Omega_{\alpha}(s) \delta_{\alpha, \ell=0}]$, where $\Omega_{\alpha}$ is the reduced energy density in the $\alpha$-th flow.  Writing out the spherical Bessel function $j_0(q) = \sin(q)/q$ explicitly, we find the 
 the total HDM density in this picture to be
\begin{equation}
f_{\rm hdm} \delta_{\rm hdm}(\vec{k},s) = -k^2 \sum_{\alpha=1}^{N_\tau} \frac{\Omega_\alpha(s)}{\Omega_{\rm m}(s)} \int_{s_{\rm i}}^s {\rm d} s' \, a^2 (s') \, (s-s')\, \Phi(\vec{k},s') \, \frac{\sin(q_\alpha)}{q_\alpha},
\label{eq:monopole}
\end{equation}
where $q_\alpha = k \tau_\alpha(s-s')/m$, and 
the ratio $\Omega_{\alpha}(s)/\Omega_{\rm m}(s)$ is time-independent in the non-relativistic limit.

Consider now the equivalent expression in the Eulerian formulation, i.e., equation \eqref{eq:dh1}, rewritten here in terms of the full solution to $\langle\delta f\rangle_{\mu}$ (equation~\eqref{eq:df_sprime}, slightly recast):
\begin{equation}
\begin{aligned}
  f_{\rm hdm}  \delta_{\rm hdm}(\vec{k},s)&=\frac{f_{\rm hdm}}{C}\int_0^\infty {\rm d}p\, p^2 \ \langle\delta f\rangle_{\mu}(\vec{k},p,s) \\
  &  = \frac{k^2}{3} \frac{f_{\rm hdm}}{C}\int_0^\infty {\rm d}p\, p^3 
   \frac{\partial \bar f}{\partial p}
	\int_{s_{\rm i}}^s \mathrm{d}s' a^2(s') (s-s') \Phi(\vec{k},s') {\cal W}(q),
 \label{eq:eulerian}
\end{aligned}
\end{equation}
with ${\cal W}(q) \equiv 3 W(q)/q = 3\sin(q)/q^3 - 3\cos(q)/q^2$, $C\equiv \int_0^\infty {\rm d} p\, p^2\ \bar{f}(p)$, and $q \equiv k p (s-s')/m$.  Integrating the second line by parts, we find
\begin{equation}
  f_{\rm hdm}  \delta_{\rm hdm}(\vec{k},s)
    = - k^2 \frac{f_{\rm hdm}}{C}\int_0^\infty {\rm d}p\, p^2  \bar{f}
	\int_{s_{\rm i}}^s \mathrm{d}s' a^2(s') (s-s') \Phi(\vec{k},s')\left[{\cal W}(q) +\frac{1}{3} \frac{\partial {\cal W}}{\partial \ln q} \right].
\end{equation}
It is straightforward to verify that 
\begin{equation}
{\cal W}(q) +\frac{1}{3} \frac{\partial {\cal W}}{\partial \ln q}  =\frac{\sin (q)}{q} = j_0(q).
\label{eq:mappingwj}
\end{equation}
Then, identifying $\tau_\alpha  \leftrightarrow p$, and 
\begin{equation}
\sum_{\alpha=1}^{N_\tau} \Omega_{\alpha}(s) \leftrightarrow   \Omega_{\rm hdm}(s)\int_0^\infty \frac{{\rm d} p\, p^2 \bar{f}}{\int_0^\infty {\rm d} p\, p^2 \bar{f}},
\end{equation}
we see that multi-fluid formulation~\eqref{eq:monopole} and the standard Eulerian formulation~\eqref{eq:eulerian} yield the same result in the limit of large $N_\tau$.  The mapping~\eqref{eq:mappingwj} between ${\cal W}(q)$ and $j_0(q)$ is the origin of the relation~\eqref{eq:FtoG} between the linear response functions ${\cal F}$ and ${\cal G}$ of the two formulations.


\section{Derivation of the SuperEasy gravitational potential}\label{app:derivations}

We begin with the Poisson equation~\eqref{eq:poisson1}, and define a modification factor $\tilde{g}(k,s)$ via
\begin{equation}
k^2 \Phi(\vec{k},s) = -\frac{3}{2} {\cal H}^2 \, \Omega_{\rm cb}(s) \, \tilde{g}(k,s) \,
\delta_{\rm cb} (\vec{k},s),
\end{equation}
where 
\begin{equation}
\tilde{g}(k,s) \equiv 1+ {\vec{f}_{\rm h}}^{~T}  
\left[\hat{I}_N+\hat{M}(k,s)  \right]^{-1}
\vec{L}(k,s),
\end{equation}
and we remind the reader that it is $\Omega_{\rm cb}(s)$, not $\Omega_{\rm m}$, that appears in the prefactor.  In the limit $k \to 0$, we require $\tilde{g} \to 1/f_{\rm cb}$, while $\tilde{g} \to 1$ as $k \to \infty$.  It is therefore more convenient to work with 
\begin{equation}
f_{\rm cb}\,\tilde{g}(k,s) \equiv \left\{1+ {\vec{f}_{\rm h}}^{~T}  
\left[\hat{I}_N+\hat{M}(k,s)  \right]^{-1}
\vec{L}(k,s) \right\} \left(1-\sum_{i=1}^N f_{{\rm h}_i}  \right),
\end{equation}
in order to recover the limiting behaviours exactly.

To obtain $f_{\rm cb}\,\tilde{g}(k,s)$ to ${\cal O}(f_{{\rm h}_{\rm i}}^2)$, we 
expand $[\hat{I}_N-\hat{M}(k,s)]^{-1}$ as per equation~\eqref{eq:noinversion} to ${\cal O}(f_{{\rm h}_{\rm i}})$, 
\begin{equation}
\begin{aligned}
f_{\rm cb}\,\tilde{g}(k,s) & \simeq  \left\{1+ {\vec{f}_{\rm h}}^{~T}  
\left[\hat{I}_N+\hat{M}(k,s) +{\cal O}(f_{{\rm h}_{\rm i}}^2)  \right]
\vec{L}(k,s) \right\} \left(1-\sum_{i=1}^N f_{{\rm h}_i}  \right)\\
& =
\left\{1+ \sum_{i=1}^N f_{{\rm h}_i} 
 L_i(k,s)\bigg[1+ \sum_{j\neq i}^N  f_{{\rm  h}_j} L_j(k,s)\bigg] +{\cal O}(f_{{\rm h}_{\rm i}}^3) \right\} \left(1-\sum_{k=1}^N f_{{\rm h}_k}  \right)\\
& \simeq 1 +  \sum_{i=1}^N f_{{\rm h}_i} 
 \left[L_i-1\right] -\sum_{i=1}^N \sum_{j=1}^N f_{{\rm h}_i} f_{{\rm h}_j} L_i
 +
\sum_{i=1}^N \sum_{j\neq i}^N f_{{\rm h}_i}  L_i  f_{{\rm  h}_j} L_j  +{\cal O}(f_{{\rm h}_{\rm i}}^3),
\label{eq:modfac1}
\end{aligned}
\end{equation}
where we have used the definition~\eqref{eq:mmatrix} of the matrix~$\hat{M}$.  Rewriting $L_i(k,s)\equiv \mathcal{G}_i(k,s)/[1-\mathcal{G}_i(k,s)f_{{\rm h}_i}]$ in terms of the interpolation function ${\cal G}_i(k,s)$, we find to ${\cal O}(f_{{\rm h}_{\rm i}}^2)$ 
\begin{equation}
f_{\rm cb}\, \tilde{g}(k,s)
 \simeq  
1 + \sum_{i=1}^N f_{{\rm h}_i}  
 \big[{\cal G}_i -1\big]+ \sum_{i,j=1}^N  f_{{\rm h}_i} f_{{\rm h}_j}  {\cal G}_i \big[{\cal G}_j-1\big]+ {\cal O}(f_{{\rm h}_{\rm i}}^3),
\end{equation}
and subsequently the Poisson equation
\begin{equation}
k^2 \Phi(\vec{k},s)
  \simeq  -\frac{3}{2} {\cal H}^2 \, \Omega_{\rm cb}(s) \frac{1}{f_{\rm cb}}\left\{ 
1 + \sum_{i=1}^N  f_{{\rm h}_i}\big[{\cal G}_i-1\big] + \sum_{i,j=1}^N  f_{{\rm h}_i} f_{{\rm h}_j}  {\cal G}_i \big[{\cal G}_j-1\big] \right\} \delta_{\rm cb} (\vec{k},s).
\end{equation}
This is the centrepiece of the generalised SuperEasy approach. The modification factor depends only on the magnitude $k \equiv |\vec{k}|$.   As $k \to 0$, ${\cal G} \to 1$, and we recover $f_{\rm cb} \,\tilde{g}= \{\cdots \} \to 1$.  Similarly, as $k \to \infty$, ${\cal G} \to 0$, from which we find $f_{\rm cb} \,\tilde{g} =\{\cdots\} \to 1-f_{\rm hdm} = f_{\rm cb}$. The choice of $N$ and the momentum-binning scheme are defined by the user, guided by convergence tests.  
Equation~\eqref{eq:gse_poisson} shows the leading-order result.

\bibliography{bib.bib}
\bibliographystyle{bibi}

\end{document}